\documentstyle[prb,aps]{revtex}
\tighten
\begin{document}

\title{The Low-Energy Fixed Points of Random Quantum Spin Chains}

\author{E. Westerberg,\cite{EWest} A. Furusaki,\cite{AFuru}
  M. Sigrist,\cite{MSigr} and P. A. Lee}

\address{Department of Physics, Massachusetts Institute of Technology,
Cambridge, Massachusetts 02139}

\maketitle

\begin{abstract}
The one-dimensional isotropic quantum Heisenberg spin systems with
random couplings and random spin sizes are investigated
using a real-space renormalization group scheme.
It is demonstrated that these systems belong to a
universality class of disordered spin systems,
characterized by weakly coupled large effective spins.
In this large-spin phase the uniform magnetic susceptibility diverges as
$T^{-1}$ with a non-universal Curie constant at low temperatures $T$,
while the specific heat vanishes as $T^\delta |\ln T|$ for $T\to 0$.
For broad range of initial distributions of couplings and spin sizes
the distribution functions approach a single fixed-point form, where
$\delta \approx 0.44$.
For some singular initial distributions, however, fixed-point
distributions have non-universal values of $\delta$, suggesting that
there is a line of fixed points.
\end{abstract}

\pacs{74.20.De, 74.50.+r, 74.72} 

%\narrowtext

\section{Introduction}
\label{sec:intr}

Over many decades one-dimensional (1D) quantum spin systems 
(`quantum spin chains') have attracted 
a lot of interest and led to the development of many theoretical methods
which are now commonly used for the study of other highly
correlated systems.\cite{frad1}
Despite the apparent simplicity of quantum spin chains, they show a wealth of
physical properties which give a key to our understanding of various
phenomena, e.g., quantum phase transitions, topological order, and
fractional statistics.\cite{frad1,romm1,hald2}
Since the discovery of various quasi-1D materials,
the study of 1D spin systems, which is mainly based on the Heisenberg
model and its variations, is also of experimental relevance.
Examples of such materials include so-called NINO,
NENP,\cite{rena1,mill1} and ${\rm Sr}_3{\rm CuPtO}_6$.\cite{wilk1}
In particular, the latter system belongs to a class of compounds which
is compositionally very flexible and has been under intense
experimental investigation over the last few years.
This type of quasi-1D system was first reported in
${\rm Sr}_4{\rm PtO}_6$ by Randall and Katz,\cite{rand1}
and it is now possible to produce compounds of the form
${\rm Sr}_3{\rm MNO}_6$ in various combinations with M = Cu, Mg, Zn,
Yb, Na, Ca, Co and N = Pt, Ir, Rh, Bi.

Disorder effects play a particularly important role in 1D quantum spin
systems, as even small deviations from the regular system often
destabilize the pure phases.\cite{doty1}
Real experimental systems naturally contain impurities
and other types of disorder.
Therefore it is very important to understand the influence of disorder
on the properties of such systems in order to interpret experimental
results.  
To our knowledge, the first 1D spin system recognized for its disorder
belongs to the class of charge-transfer salts TCNQ
(tetracyanoquinodimethanide).\cite{bula1}
These systems have been successfully described by a 1D spin-$\frac{1}{2}$
Heisenberg model with random strength of antiferromagnetic
exchange couplings between the spins.
A more recent example of disordered spin chains is
${\rm Sr}_3{\rm CuPt}_{1-x}{\rm Ir}_x{\rm O}_6$.\cite{nguy1}
While the pure compounds ${\rm Sr}_3{\rm CuPtO}_6$ ($x=0$) and
${\rm Sr}_3{\rm CuIrO}_6$ ($x=1$) are antiferromagnetic (AF) and
ferromagnetic (FM) spin systems, respectively, the alloy
${\rm Sr}_3{\rm CuPt}_{1-x}{\rm Ir}_x{\rm O}_6$ contains both AF and
FM couplings.
The fraction of FM bonds is simply related to $x$, the concentration
of Ir ions.
In a previous work, we modeled
${\rm Sr}_3{\rm CuPt}_{1-x}{\rm Ir}_x{\rm O}_6$ with a
nearest-neighbor spin-$\frac{1}{2}$ Heisenberg chain, where the
exchange coupling between neighboring spins is $+J$ or $-J$ with
probability $p$ and $1-p$ respectively.
The methods we used (high-temperature expansion and
transfer matrix approximation) give reliable results down to
$k_BT\sim J/5$, where $T$ is the temperature and $k_B$ is Boltzmann's
constant.
In this regime the numerically calculated magnetic susceptibility
$\chi(T)$ and the specific heat $C(T)$ interpolate smoothly between the
corresponding quantities of purely FM ($p=1$) and AF ($p=0$) chains,
giving good qualitative agreement with experimental data.\cite{furu1,furu2}
At temperatures below $\sim J/k_B$ the effects of disorder become
significant.
We demonstrated that in this temperature regime spins correlate within
AF and FM segments of the chain separately.
The emerging new degrees of freedom which dominate the
thermodynamics are (large) effective spins each corresponding to a
correlated segment.
The size of these spins and their residual interaction
are set by the local disorder and hence is random.
From exact diagonalization of finite segments we concluded that the
low-temperature physics of the random spin system is described by the
effective Hamiltonian,
\begin{equation}
{\cal H} = \sum_i J_i{\bf S}_i\cdot{\bf S}_{i+1} \ \ , 
\label{eq:ham1}
\end{equation}
where both the couplings $J_i$, which may have either sign, and the
spin sizes $S_i$ are random.
In particular, we emphasize that the resulting distribution of $J_i$
in Eq.~(\ref{eq:ham1}) is broad and dense, in contrast to the discrete 
distribution of the initial model.
In this paper we take Eq.~(\ref{eq:ham1}) as starting point.
We discuss the low-temperature properties of this model, and the
various fixed points encountered for different initial distributions of
couplings and spins.\cite{sigfuru}

Before going into details we briefly summarize our results and the
method we use, which is a generalization of the RSRG scheme introduced
by Ma, Dasgupta and Hu (MDH) in 1979 to study the 1D
spin-$\frac{1}{2}$ random antiferromagnet (RAF).\cite{ma1}
The RAF, where all couplings are antiferromagnetic but vary in
magnitude, has also been investigated using Kadanoff block spin RG
techniques,\cite{hirs1} and more recently by the density
matrix RG method.\cite{hida2}
The method of Dasgupta {\it et al.} has proven to be the most
successful one, and was recently extended by Fisher,\cite{fish1}
who solved the RG equations exactly. 

In the MDH RSRG scheme, a decimation of degrees of freedom occurs
through the successive formation of spin singlets from the most
strongly coupled spin pairs.
This scheme conserves the form of the Hamiltonian in the original model, 
but changes the distribution of couplings, which gradually
approaches a fixed-point form.
The model in Eq.~(\ref{eq:ham1}) contains arbitrary spin sizes and
couplings with
random sign, so that, in general, two correlated spins do not
combine into a singlet.
Rather they form an effective spin with renormalized
couplings to its neighbors.
Here we introduce a modified RSRG scheme which takes this into
account.
Like the MDH RSRG scheme it conserves the form of the Hamiltonian,
Eq.~(\ref{eq:ham1}), but changes the distributions of couplings and gaps.
The RG flow generated can therefore be thought of
as a flow in the space of distributions of couplings and gaps.\cite{west1}
We demonstrate that for a wide range of initial distributions of
couplings and spins, the RG flow of the distribution functions eventually
approaches a single universal fixed point.
This fixed point represents the following physical properties.  
Both entropy and specific heat vanish as $T^{2\alpha}|\ln T|$, where
the power $\alpha\approx 0.22$ is rather small.
The exponent $\alpha$ appears also in the non-linear magnetization
where $M(H) \propto H^{\alpha/ 1 + \alpha}$ for sufficiently large
fields $H$.
For very singular initial distribution of the couplings we find that 
$\alpha$ takes a non-universal value, suggesting 
the presence of a fixed line.
We also find a surprising fact that, for both the universal and
non-universal fixed points, the susceptibility follows the Curie
behavior down to zero temperature with a non-universal Curie constant.
We finally show how the two previously known random phases, the random
singlet phase (RSP) and the random dimer solid
(RDS)\cite{hyma1,hyma2}, both of which correspond to Eq.~(\ref{eq:ham1}) 
with all $S_i=1/2$ and all $J_i>0$, are unstable against the admixture
of an arbitrarily small concentration of FM couplings and/or larger spins. 

Our paper has the following structure.
We start with a brief review of the MDH RSRG scheme in
Sec.~\ref{sec:trgs1}, before we generalize it to chains with both AF
and FM couplings in Sec.~\ref{sec:trgs2}.
In Sec.~\ref{sec:scal} we analyze the distributions of spins
and couplings, and their scaling forms close to a fixed point.
We perform the RG scheme numerically by simulating random spin
chains with various initial distributions of couplings and spins.
The numerical results shown in Sec.~\ref{sec:nume} confirm the scaling
forms conjectured in Sec.~\ref{sec:scal}, but also reveal that random
spin chains with very singular initial distributions of gaps flow to
non-universal fixed points.
In Sec.~\ref{sec:ther} we derive the scaling forms of thermodynamic
quantities, and in Sec.~\ref{sec:comm} we comment on the
approximations involved in the RG transformation.
Finally we summarize our results in Sec.~\ref{sec:conc}, and
compare the large-spin phase (LSP) to other disordered phases.
We also discuss the stability of the various phases and in particular
the RG flow between the LSP, RSP, and RDS.

\section{The renormalization-group scheme}
\label{sec:trgs}
To study the low-temperature properties of systems described by
Eq.~(\ref{eq:ham1}), we generalize a RSRG method
introduced by Ma, Dasgupta and Hu (MDH).\cite{ma1}
We start with a brief
review of the MDH scheme for the RAF with $ S_i =1/2 $ and random $ J_i >0 $.

\subsection{The MDH RG for antiferromagnetic spin-$\frac12$ chains}
\label{sec:trgs1}
Consider an antiferromagnetic nearest-neighbor Heisenberg spin-$\frac12$
chain in which the largest coupling is $J_0$ and the remaining 
couplings $J_i$ are distributed according to
$P(J_0;J_i)$.
We focus on the link with the largest coupling,
$J_i=J_0$, and the terms in the Hamiltonian
(\ref{eq:ham1}) that involve the spins ${\bf S}_i$ and
${\bf S}_{i+1}$, (see Fig.~\ref{fig:dama}a),
\begin{equation}
{\cal H}^\prime = {\cal H}^\prime_0 +{\cal H}^\prime_I
\label{eq:ham2}
\end{equation}
where
\begin{eqnarray*}
{\cal H}^\prime_0 & = & J_0{\bf S}_i\cdot{\bf S}_{i+1} \ \ , \\
{\cal H}^\prime_I & = & J_{i-1}{\bf S}_{i-1}\cdot{\bf S}_{i}+
                        J_{i+1}{\bf S}_{i+1}\cdot{\bf S}_{i+2} \ \ .
\end{eqnarray*}
If the distribution $P(J_0;J)$ is broad, $J_{i\pm 1}$ are
typically much smaller than $J_0$ and we can treat ${\cal H}^\prime_I$
as a perturbation to ${\cal H}^\prime_0$. In the ground state of
${\cal H}^\prime_0$ the spins ${\bf S}_i$ and ${\bf S}_{i+1}$ form a
singlet, and the energy gap to the excited states is $J_0$. 
This ground state is four-fold degenerate
since the unperturbed Hamiltonian  ${\cal H}^\prime_0$ does not involve the 
directions of ${\bf S}_{i-1}$ and ${\bf S}_{i+2}$.
${\cal H}^\prime_I$ lifts the degeneracy and splits the unperturbed
ground state into a singlet and a triplet, and the low energy spectrum
of the four-spin Hamiltonian (\ref{eq:ham2}) is described by an
effective Hamiltonian 
\begin{equation}
{\cal H}^{\rm eff} = \widetilde{J}{\bf S}_{i-1}\cdot{\bf S}_{i+2} \ \ .
\label{eq:ham3}
\end{equation}
The effective coupling $\widetilde{J}$ is determined from the energy
splitting of the unperturbed ground state, and to second order in 
$J_{i\pm 1}/J_0$
the result is $\widetilde{J}/J_0=\frac{J_{i-1}J_{i+1}}{2J_0^2}$.
Physically the weak interaction between ${\bf S}_{i-1}$ and
${\bf S}_{i+2}$ is mediated by exciting virtual triplet  states in the
interjacent spin pair. 

In the Hamiltonian (\ref{eq:ham1}) we replace the terms in ${\cal H}^\prime$
with the effective interaction  ${\cal H}^{\rm eff}$ in
Eq.~(\ref{eq:ham3}) to get an effective Hamiltonian for the low-energy
degrees of freedom of the spin chain.
Repeating this procedure and successively replacing the strongest
remaining coupling in the chain preserves the form of the Hamiltonian
but changes the distribution of couplings and, in particular, lowers
$J_0$, the largest remaining coupling in the chain.
If $P(J_0^\prime ,J)$ is the distribution of couplings at a point when
the largest remaining coupling is $J_0^\prime$, then the removal of
bonds $J_i\in [J_0-dJ_0,J_0]$ generates a flow equation
for $P(J_0,J)$ \cite{ma1}
\begin{equation}
\frac{dP(J_0;J)}{dJ_0} = -P(J_0;J_0)
\int_0^{J_0}\!\!dJ_1dJ_2P(J_0;J_1)P(J_0;J_2)\delta (J-J_1J_2/2J_0) \ \ .
\label{eq:flow1}
\end{equation}
The flow equation (\ref{eq:flow1}) is derived under the assumption
that there are no spatial correlations among the bond strengths.
This is indeed the case if there are no correlations in the distribution
of couplings in the initial chain. 
It has been shown that if the initial distribution
of bonds is normalizable, Eq.~(\ref{eq:flow1})
has a unique fixed-point solution that governs the low-energy physics
of random bond antiferromagnetic spin-$\frac12$ chains.\cite{fish1,foot1}

\subsection{Generalization of the MDH RG}
\label{sec:trgs2}
We apply the same strategy to the random spin chain with couplings of
either sign and random spin sizes.
In contrast to the previous case, a link is determined not only by the 
coupling strength but also by its left and right spin, 
$\{\Delta_i ,S_i,S_{i+1}\}$ (see Fig.~\ref{fig:dama}b).
We define $\Delta_i$ as the energy gap between the ground state 
multiplet and the first excited multiplet in the corresponding
two-spin Hamiltonian  ${\cal H}=J_i{\bf S}_i\cdot{\bf S}_{i+1}$:
\cite{note-hyman}
\begin{equation}
\Delta_i =  \Bigg\{ \begin{array}{llll}
  |J_i|(S_i+S_{i+1}) & : & J_i<0 & {\rm (ferromagnetic \ link)}\ \ , \\
  J_i(|S_i-S_{i+1}|+1) & : & J_i>0 & {\rm (antiferromagnetic \ link)\ \ .} 
\end{array}
\label{eq:gaps}
\end{equation}
We assume a broad distribution of interaction energies and focus
on the link in the chain which, if completely isolated from the rest of the
chain, requires the largest energy $\Delta =\Delta_0$ to excite the ground
state multiplet.
We consider the situation illustrated in Fig.~\ref{fig:dama}c, where
$\{\Delta_0,S_L,S_R\}$ is the strongest link in the chain, and
$\{\Delta_1,S_1,S_L\}$ and $\{\Delta_2,S_R,S_2\}$ are its adjacent links.
In the spirit of the MDH RSRG scheme we replace the strongest link 
$\{\Delta_0,S_L,S_R\}$ with an effective spin of size $S=|S_L\pm S_R|$
representing the lowest-energy multiplet of
maximum (minimum) spin for a ferromagnetic (antiferromagnetic) link.
The residual effective interaction for ${\bf S}_{1}$, ${\bf S}$, and 
${\bf S}_{2}$ is calculated perturbatively in 
$\varepsilon_{1,2}=\Delta_{1,2}/\Delta_0$.
The effective interaction is isotropic, and to first order
in $\varepsilon_{1,2}$ given by
\begin{equation}
{\cal H}^{\rm eff} = \widetilde{J}_1{\bf S}_1\cdot{\bf S}+
\widetilde{J}_2{\bf S}\cdot{\bf S}_2
\end{equation}
with
\begin{mathletters}
\label{eq:egap}
\begin{eqnarray} 
J_0>0 \ , \ J_1>0 \ , \ S_L>S_R & \ \ \Longrightarrow \ \ & 
\widetilde{\Delta}_1 =\Delta_1f_1(S_1,S_L,S_R) \ ; \ \widetilde{J}_1>0 \\
J_0>0 \ , \ J_1>0 \ , \ S_L<S_R & \ \ \Longrightarrow \ \ & 
\widetilde{\Delta}_1 =\Delta_1f_2(S_1,S_L,S_R) \ ; \ \widetilde{J}_1<0 \\
J_0>0 \ , \ J_1<0 \ , \ S_L>S_R & \ \ \Longrightarrow \ \ & 
\widetilde{\Delta}_1 =\Delta_1f_3(S_1,S_L,S_R) \ ; \ \widetilde{J}_1<0 \\
J_0>0 \ , \ J_1<0 \ , \ S_L<S_R & \ \ \Longrightarrow \ \ & 
\widetilde{\Delta}_1 =\Delta_1f_4(S_1,S_L,S_R) \ ; \ \widetilde{J}_1>0 \\
J_0<0 \ , \ J_1>0 \qquad \qquad \quad & \ \ \Longrightarrow \ \ &
\widetilde{\Delta}_1 = \Delta_1f_5(S_1,S_L,S_R) \ ; \ \widetilde{J}_1>0 \\ 
J_0<0 \ , \ J_1<0 \qquad \qquad \quad & \ \ \Longrightarrow \ \ &
\widetilde{\Delta}_1 = \Delta_1f_6(S_1,S_L,S_R) \ ; \ \widetilde{J}_1<0
\end{eqnarray}
\end{mathletters}\noindent
where
\begin{mathletters}
\label{eq:traf}
\begin{eqnarray}
f_1(S_1,S_L,S_R) & = & 
\frac{(S_L+1)(|S_1-S_L+S_R|+1)}{(S_L-S_R+1)(|S_1-S_L|+1)}  \\
f_2(S_1,S_L,S_R) & = & 
\frac{S_L(S_1+S_R-S_L)}{(S_R-S_L+1)(|S_1-S_L|+1)}  \\
f_3(S_1,S_L,S_R) & = & 
\frac{(S_L+1)(S_1+S_L-S_R)}{(S_L-S_R+1)(S_1+S_L)}  \\
f_4(S_1,S_L,S_R) & = & 
\frac{S_L(|S_1-S_R+S_L|+1)}{(S_R-S_L+1)(S_1+S_L)}  \\
f_5(S_1,S_L,S_R) & = & 
\frac{S_L(|S_1-S_L-S_R|+1)}{(S_L+S_R)(|S_1-S_L|+1)}  \\
f_6(S_1,S_L,S_R) & = & 
\frac{S_L(S_1+S_L+S_R)}{(S_L+S_R)(S_1+S_L)} \ \ .
\end{eqnarray}
\end{mathletters}\noindent
A derivation of these equations is shown in  Appendix~\ref{sec:appgapA}.
From the knowledge of the gap $\widetilde{\Delta}_1$ and the sign of 
$\widetilde{J}_1$, $\widetilde{J}_1$ is readily calculated via
Eq.~(\ref{eq:gaps}).
Similarly $\widetilde{\Delta}_2$ is obtained by replacing 
$S_1 $ by $ S_2$ and $S_L $ by $ S_R$ in 
Eqs.~(\ref{eq:egap}) and (\ref{eq:traf}).
These equations do not require the spins to be multiples of $1/2$, and
from now on we regard spins as continuous variables. 
The case where the strongest link is antiferromagnetic with $S_L=S_R$
is not accounted for in Eq.~(\ref{eq:egap}). 
In this case the two spins $S_L$ and $S_R$ form a singlet, and the
leading order contribution to the effective coupling between $S_1$ and $S_2$
is (c.f. Appendix~\ref{sec:appgapB})
\begin{equation}
\widetilde{J} = \frac{2J_1J_2}{3J_0}S_L(S_L+1) \ \ ,
\label{eq:seco}
\end{equation}
which is easily translated into $\tilde\Delta$ via Eq.~(\ref{eq:gaps}). 
For $S_L=S_R=1/2$ this is the original MDH RG transformation.

As in the original RSRG scheme, the effect of successively forming
effective spins is to change the distributions of gaps and spins (links)
without changing the form of the Hamiltonian.
In analogy to the probability distribution for the couplings in the
RAF, we define probability distributions of ferromagnetic ($ P^F $) 
and antiferromagnetic ($ P^A $) links where the largest remaining gap
in the chain is $\Delta_0$,
\begin{mathletters}
\label{eq:pros1}
\begin{eqnarray}
 & P^F(\Delta_0;\Delta,S_L,S_R) & \ \ , 
\label{eq:prof1} \\
 & P^A(\Delta_0;\Delta,S_L,S_R) & \ \ .
\label{eq:proa1}
\end{eqnarray}
\end{mathletters}\noindent
The probability distributions $P^{A,F}$ are symmetric in
$S_L$ and $S_R$ and obey the normalization condition 
\begin{equation}
\int_0^{\Delta_0} d\Delta \int_0^\infty dS_LdS_R \left[
P^F(\Delta_0;\Delta,S_L,S_R)+P^A(\Delta_0;\Delta,S_L,S_R)\right] = 1 \ \ ,
\label{eq:norm}
\end{equation}
for any value of $\Delta_0$.
From $P^{A}$ and
$P^F$ we can calculate the distributions of spins, gaps and coupling 
constants. As a special case, the original spin-$\frac12$
antiferromagnetic chain studied in Ref.~\onlinecite{ma1} corresponds to 
\begin{mathletters}
\label{eq:heid}
\begin{eqnarray}
P^A(\Delta_0;\Delta,S_L,S_R) & = & 
\delta (S_L-\textstyle{\frac12})
\delta (S_R-\textstyle{\frac12})P(\Delta_0;\Delta)  \ \ , \\
P^F(\Delta_0;\Delta,S_L,S_R) & = & 0 \ \ .
\end{eqnarray}
\end{mathletters}\noindent
If there are no correlations between neighboring links (except for the
obvious correlation that they share one spin), the flow equations for
$P^{A,F}$ are
\begin{mathletters}
\label{eq:flow2}
\begin{eqnarray}
\frac{dP^A}{d\Delta_0} & = & F_1[P^A,P^F]\ \ ,  \\
\frac{dP^F}{d\Delta_0} & = & F_2[P^A,P^F]\ \ ,
\end{eqnarray}
\end{mathletters}\noindent
which generalize the MDH RSRG flow equation (\ref{eq:flow1}).
In Eqs.~(\ref{eq:flow2}) $F_1$ and $F_2$ are two (non-linear) functionals
of $P^A$ and $P^F$, whose explicit forms depend on the functions $f_n$ in
Eq.~(\ref{eq:traf}), c.f. Appendix~\ref{sec:appflow}.

\section{Scaling forms of the probability distributions}
\label{sec:scal}
As links are replaced by effective spins, the effective couplings and,
hence, the gaps of the links decrease.
At the same time the average distance $na_0$ between neighboring
effective spins as well as the magnitude of the effective spins
increase.
(Here $a_0$ is the original lattice constant and $n$ is the ratio
of the number of original spins to the number of effective spins.)
We expect that the link distributions 
eventually approach a fixed point where $P^A$, $P^F$ and $n$
exhibit scaling behavior
\begin{mathletters}
\label{eq:scas1}
\begin{eqnarray}
P^A(\Delta_0;\Delta,S_L,S_R) & = & \Delta_0^{-\gamma_A}
Q^A\!\left(\frac{\Delta}{\Delta_0^{\beta_A}},
           \frac{S_L}{\Delta_0^{-\alpha_A}},
           \frac{S_R}{\Delta_0^{-\alpha_A}}\right) \ \ ,
\label{eq:scaa1}  \\
P^F(\Delta_0;\Delta,S_L,S_R) & = & \Delta_0^{-\gamma_F}
Q^F\!\left(\frac{\Delta}{\Delta_0^{\beta_F}},
           \frac{S_L}{\Delta_0^{-\alpha_F}},
           \frac{S_R}{\Delta_0^{-\alpha_F}}\right) \ \ ,
\end{eqnarray}
\label{eq:scaf1}
\end{mathletters}\noindent
and
\begin{equation}
n \sim \Delta_0^{-\delta} \ \ .
\label{eq:scal1}
\end{equation}
The exponent $ \delta $ is related to the dynamical exponent $ z $ ($ z =
1/\delta$). The seven exponents in Eqs.~(\ref{eq:scas1}) and (\ref{eq:scal1})
are not all independent. Indeed, we argue below that
\begin{mathletters}
\label{eq:exres}
\begin{eqnarray}
\alpha_F = \alpha_A & \equiv & \alpha 
\label{eq:exre1} \\
\beta_F = \beta_A & = & 1 
\label{eq:exre2} \\
\gamma_F = \gamma_A & = & 1-2\alpha 
\label{eq:exre3} \\
\delta & = & 2\alpha 
\label{eq:exre4}
\end{eqnarray}
\end{mathletters}\noindent
so that in the scaling regime  
\begin{mathletters}
\label{eq:scas2}
\begin{eqnarray}
P^A & = & \frac{1}{\Delta_0^{1-2\alpha}}Q^A
          (\Delta /\Delta_0,S_L\Delta_0^\alpha ,S_R\Delta_0^\alpha ) 
\label{eq:scaa2} \\
P^F & = & \frac{1}{\Delta_0^{1-2\alpha}}Q^F
          (\Delta /\Delta_0,S_L\Delta_0^\alpha ,S_R\Delta_0^\alpha )
\label{eq:scaf2}
\end{eqnarray}
\end{mathletters}\noindent
with length scaling as
\begin{equation}
n \sim \Delta_0^{-2\alpha} \ \ .
\label{eq:scal2}
\end{equation}
The relations (\ref{eq:exres}) have been 
confirmed in numerical simulations ({\it c.f.} Sec.~\ref{sec:nume})
and can be understood as follows.
 
Let $x=N_A/(N_A+N_F)$ be the fraction of AF links in the chain.
Both FM ($x=0$) and AF ($x=1$) chains are unstable towards a small
concentration of couplings of the opposite sign.
To see that $x=1$ is unstable, we note that unless the effective
spin formed is a singlet, the removal of a link in an AF chain converts 
one neighboring link into a FM link.
Similarly $x=0$ is unstable because an
isolated AF link in a FM environment always
survives (the removal of a FM link does not change the signs of its
neighboring links, and if the AF link itself is removed, one of
its FM neighbors is converted into an AF link). 
This implies that for small enough $x$ the absolute number of AF
links is constant, so that the fraction  $x$ of AF links
increases as links are removed.
Thus, unless we start with a completely FM random spin chain or a
purely AF random spin chain with uniform magnitude of the spins, the
fixed-point distribution contains both FM and AF links. 
Having established that both $x_0$ and $1-x_0$, the fraction of AF and
FM links at the fixed point respectively, are non-zero, we can easily
derive the relations (\ref{eq:exre1}-c).
Since a finite fraction of the spins belong to both a FM and an AF
link, there cannot be a separation in scales between spin sizes in AF
and FM links, i.e., $\alpha_A=\alpha_F=\alpha$.
$\beta_F=1$ ($\beta_A=1$) follows from the fact that the average gap
in the FM (AF) distribution, when measured in units of $\Delta_0$
is finite and independent of $\Delta_0$.
$\gamma_A=1-2\alpha$ follows trivially from the finiteness of $x_0$
\begin{eqnarray}
x_0 & = &
\int_0^{\Delta_0}d\Delta\int_0^\infty dS_L dS_R
\Delta_0^{-\gamma_A}Q^A(\Delta /\Delta_0,\Delta_0^\alpha S_L,
\Delta_0^\alpha S_R) \nonumber \\
 & = & \Delta_0^{1-2\alpha -\gamma_A}
\int_0^1 d\Delta^\prime\int_0^\infty ds^\prime ds^{\prime\prime}
Q^A(\Delta^\prime ,s^\prime ,s^{\prime\prime}) \ \ ,
\label{eq:normA}
\end{eqnarray}
which must be independent of $\Delta_0$.
An analogous argument for $1-x_0$ implies  $\gamma_F=1-2\alpha$.

The last relation (16d) requires a more detailed analysis of the way
effective spins are formed by correlating the original spins in
clusters.
The size of the effective spin corresponds to the spin
quantum number of the ground state of the cluster.
Since the spin system is not frustrated the spin quantum number is
determined from the classical correlation of the spins
(parallel and antiparallel).
The total spin of a cluster of $n$ spins is then given by 
\begin{equation}
S = \left| \displaystyle\sum_{i=1}^n \pm S_i\right| \ \ ,
\label{eq:rawa1}
\end{equation}
where two neighboring spins enter the sum with the same (opposite)
sign if their mutual coupling is ferromagnetic (antiferromagnetic).
This leads to a typical random walk problem which results
in the scaling
\begin{equation}
S \sim n^{1/2} \ \ .
\label{eq:rawa2}
\end{equation}
From this we conclude $ \delta = 2 \alpha $.

The remaining independent exponent, which we take to be $\alpha$, is
related to the average renormalization of the gaps.
Consider two neighboring effective spins
with a coupling corresponding to  a gap $\Delta$.
Suppose also that each of the two effective spins is made up of $2^k$
spins at some larger energy scale $\Delta^\prime$.
As the energy scale is lowered from $\Delta^\prime$ to
$\Delta$ and the $2^k$ spins form one large effective spin, the gap
$\Delta$ is typically renormalized $2k$ times ($k$ times in the course
of formation of the left effective spin and similarly $k$ times from
the right).
If the magnitude of a gap is reduced on average by a factor $a$ each
time a neighboring link is replaced with an effective spin, then 
\begin{equation} 
n\sim n^\prime 2^k = 
n^\prime \left( a^{2k}\right)^\frac{\ln 2}{2\ln a} \sim
n^\prime \left( \frac{\Delta}{\Delta^\prime}
\right)^\frac{\ln 2}{2\ln a} \ \ ,
\label{eq:exal}
\end{equation}
from which we read off the relation
\begin{equation}
\alpha = -\frac{\ln 2}{4\ln a} \ \ .
\label{eq:reaa}
\end{equation}

Using the scaling form, we can get some information on long-distance
behavior of spin-spin correlation functions.
Let us introduce two kinds of spin-spin correlation functions which
characterizes the correlation between spins at low temperatures.
The first one is the usual spin-spin correlation function,
\begin{equation}
C_1(i-j)=\langle{\bf S}_i\cdot{\bf S}_j\rangle,
\label{eq:C_1}
\end{equation}
where $\langle\;\rangle$ represents both thermal average and average
over random configurations.
Since the number of AF bonds between ${\bf S}_i$ and ${\bf S}_j$ is
random, the two spins may either be in parallel or antiparallel.
Thus, after taking the random average the correlation function decays
exponentially for large $|i-j|$ even at zero temperature:
\begin{equation}
C_1(i-j)\propto
\exp\left\{-|i-j|\left[\ln\!\left(\frac{1}{|2p-1|}\right)
                       +i\pi\Theta(1-2p)\right]\right\},
\end{equation}
where $p$ is the density of FM bonds.
Therefore this correlation function does not reflect the correlations
leading to the formation of effective large spins.
An appropriate correlation function is
\begin{equation}
C_2(i-j)=\langle\eta_{ij}{\bf S}_i\cdot{\bf S}_j\rangle,
\label{C_2}
\end{equation}
where $\eta_{ij}=\prod^{j-1}_{k=i}{\rm sgn}(-J_k)$ for $j>i$.
At finite temperature this correlation function should also decay
exponentially for large $|i-j|$ with the correlation length $n$:
\begin{equation}
C_2(i-j)\propto\exp(-|i-j|/n)\qquad{\rm for~}T>0.
\label{C_2-T>0}
\end{equation}
From Eq.~(\ref{eq:scal2}) we find that the correlation length grows
with decreasing temperature as $n\propto T^{-2\alpha}$.
It is likely that at zero temperature the correlation
function decays algebraically as
\begin{equation}
C_2(i-j)\propto\frac{1}{|i-j|^\nu}.
\label{C_2-T=0}
\end{equation}
Unfortunately we cannot determine the value of this new
exponent $\nu$ from our numerical RG scheme.

\section{Numerical results}
\label{sec:nume}
We perform our RG scheme by numerical simulations. 
We start each simulation by generating a chain according to independent
probability distributions for gaps and spins. 
In each decimation step we pick up the strongest link in the chain,
replace it with an appropriate effective spin, and renormalize the
neighboring bonds.
To keep the number of links fixed, one site is finally added in one
end of the chain.
This procedure is then iterated until the shape of the distribution
function of links no longer changes.
In this way we have iterated sixteen chains with both non-singular and
singular initial distribution, see Tab.~\ref{tab:chains}. 

In all our simulations the distributions of links eventually converged
to some fixed-point distributions.
The distributions rather quickly take the rough forms of the fixed-point
distributions, while the final approach and, in particular, the
convergence of the exponents in Eqs.~(\ref{eq:exres}) to their final
values are very slow and take place over up to five orders of
magnitudes in length (ten orders of magnitude in energy).
Our numerical simulations demonstrate that unless the initial
distribution has a high degree of singularity for small gaps, the
distribution of links in the chain eventually flows to a universal
fixed-point distribution of AF and FM links.  
If the initial distribution of gaps is more singular than
$P(\Delta)\sim\Delta^{-y_c}$, $y_c\approx 0.7$, our numerical simulations
suggest that the corresponding fixed-point distribution is non-universal.
This is analogous to the RAF where it has been shown that extremely
singular components in the gap distribution are conserved in the RG
flow.\cite{fish1}
In the case of the RAF $y_c=1$, so that the condition for a chain to
flow to the universal fixed-point distribution coincides with
normalizability.
In contrast to the RAF, there may be physical situations where a
random AF/FM spin chain flows to a non-universal fixed-point
distribution,\cite{hyma2} c.f.\  Sec.~\ref{sec:conc}.
Below we summarize the numerical results in the case of regular and
singular gap-distributions.

\subsection{Regular and weakly singular distributions}
\label{sec:regu}
If the initial distribution of gaps is regular or at least less
singular than $P(\Delta)\sim\Delta^{-0.7}$, we find that the
link-distributions in all chains we have studied (chains A, B, C, D,
F, H, and I  in Tab.~\ref{tab:chains}) eventually converge
to the same universal distribution with the characteristic scaling
form (\ref{eq:scas2}) proposed in section \ref{sec:scal}. 
The fixed-point distribution functions are illustrated by various
cross-sections in Figs.~\ref{fig:fpdi1} and \ref{fig:fpdi2}.
We find that the ratio of AF links stabilizes around $x=0.63$, thus
confirming the conjecture in section \ref{sec:scal} that both $x_0$
and $1-x_0$ are non-zero.
The exponents $\alpha_{A,F}$ and $\beta_{A,F}$ in
Eqs.~(\ref{eq:scas1}) are deduced from the scaling of the averages
$\langle S\rangle$ and $\langle\Delta\rangle$ with $\Delta_0$ in the
scaling regime.
This is illustrated in Fig.~\ref{fig:aver} for chain C in
Tab.~\ref{tab:chains},
where the average gap and spin size are plotted versus $\Delta_0$ in a
log-log plot.
Similarly the ratio $n$ of the number of original spins to the number
of effective spins is plotted versus the maximum gap $\Delta_0$ in
Fig.~\ref{fig:aver}b, and the evolution of the exponents as functions
of $\Delta_0$ are plotted in Figs.~\ref{fig:expo}a and \ref{fig:expo}b.
We find the fixed-point values of the exponents to be
\begin{eqnarray}
\alpha_{A,F} & = & 0.22\pm 0.01 \ \ , \nonumber \\
\beta_{A,F} & = & 1.00\pm 0.005 \ \ , \nonumber \\
\delta & = & 0.44 \pm 0.02 \ \ . \nonumber
\end{eqnarray}
The exponents in the FM distribution agree with
the ones in the AF distribution within numerical accuracy.
Thus our numerical findings confirm
the scaling forms (\ref{eq:scas2}) and
(\ref{eq:scal2}) for the probability distributions of the links
as well as the relations between the exponents, Eqs.~(\ref{eq:exres}).
Identifying $\alpha$ with either
$\alpha_A$, $\alpha_F$ or $\frac12\delta$ gives consistently
\begin{equation}
\alpha = 0.22\pm 0.01 \ \ .
\label{eq:alph}
\end{equation}
An interesting observation we have made is that the ratio $2\alpha /\delta$ 
generally stabilizes to its fixed-point value of $1$ before the two 
exponents $\delta$ and $\alpha$ separately converge to their
corresponding fixed-point values.
This confirms the robustness of the `random walk' argument in
Sec.~\ref{sec:scal} leading to the relation in Eq.~(\ref{eq:exre4}).

The typical expansion parameters in the perturbative calculation  of the
renormalized gaps are the median  ratios between $\Delta^{A,F}$ and
$\Delta_0$.
At the fixed point these ratios are 
\begin{eqnarray*}
\Delta^A_t / \Delta_0 & \sim & 0.2 \ \ , \\
\Delta^F_t / \Delta_0 & \sim & 0.3 \ \ ,
\end{eqnarray*}
where we denote the median gap in the
distributions of antiferromagnetic and ferromagnetic links 
by $\Delta^{A}_t$ and $\Delta^{F}_t$ respectively.
As expected from the increase in effective spin size, the formation of
a singlet on a link ($ S_R = S_L $) becomes increasingly rare
as the fixed point is approached.

\subsection{Strongly singular distributions}
For chains with a very singular initial distribution of gaps  
the convergence to some fixed-point distribution is generally even
slower than for regular (or modestly singular) chains.
Furthermore, in these cases the fixed-point distribution that is
eventually approached as well as the scaling exponent appears to be
non-universal.
This is illustrated in Fig.~\ref{fig:sifi}, where
the fixed-point distributions for four chains are plotted.
The dotted distributions correspond to chains where the initial
gap-distributions are
$P(\Delta )\propto\Delta^{-x}$ with $x=\{ 3/4,4/5,7/8\}$
(Tab.~\ref{tab:chains}).
The solid curve is the corresponding fixed-point distribution
for regular chains discussed in Sec.~\ref{sec:regu}.
For reasons discussed below, the actual form of the distributions
corresponding to singular initial distributions may be quantitatively
incorrect, but they deviate clearly from the fixed-point distribution
of regular chains. 
Numerically we find that chains with initial gap-distribution more
singular than
\begin{equation}
P(\Delta )\sim\Delta^{-y_c}  \qquad {\rm with} \qquad
 0.65\lesssim y_c\lesssim 0.75
\label{eq:crsi}
\end{equation}
flow to non-universal fixed-point distributions.
From log-log plots of the universal fixed point-distribution of gaps,
Fig.~\ref{fig:fpdilo}, we find that the distribution of FM gaps
diverges as $P^F(\Delta )\sim\Delta^{-0.44}$ and the distribution of
AF gaps as $P^A(\Delta )\sim\Delta^{-0.70}$ for small gaps.
Thus, even from a regular link-distribution the RG transformation itself
produces a singular fixed-point distribution of gaps, where the degree
of the singularity, $P(\Delta )\sim\Delta^{-y_c}$, is set by details
in the RG transformation rather than by the initial conditions.
If a distribution is more singular than $\Delta^{-y_c}$, the singular
component is conserved in the RG flow.
This is supported by our numerical results that $y_c$ in
Eq.~(\ref{eq:crsi}) is somewhere between $0.65$ and $0.75$, which
agrees with the singularity in the universal fixed-point distribution
of gaps  $P(\Delta)\sim\Delta^{-0.70}$.
We conjecture that in chains with gap-distributions more
singular than $\Delta^{-y_c}$, the low-energy fixed point is not
determined by the RG transformation alone, but also by the singular
distribution of extremely weak links.
In these chains we expect the fixed-point distribution as well as the
value of the scaling exponent to be non-universal.

The picture we present is in close analogy to the RAF where
Fisher has shown that the flow equation (\ref{eq:flow1}) conserves
very singular components of $P(\Delta_0,\Delta )$.\cite{fish1}
In the case of the RAF $y_c=1$, implying that any normalizable (and
hence physical) initial distribution eventually flows to a universal
fixed-point distribution. 

The fact that a very singular fixed-point distribution is dominated by
the weakest links in the initial chain casts some doubt on the
numerical results for such chains at low energies.
Indeed, since all our chains have only a finite number of links, the
singularity cannot be resolved perfectly.
Hence the number of initially extremely weak links which are
important for the low-energy behavior of the chain are relatively few
and we do not expect the numerical results for the extremely singular
chains to be quantitatively correct.
This explains why the singularity seems to soften at very low energies
while we argue that it should remain constant.
Still, the qualitative result that regular (or slightly singular)
chains flow to a universal fixed-point distribution while more
singular distributions do not, should be correct.

\section{Thermodynamics}
\label{sec:ther}
\subsection{Entropy and specific heat}
\label{sec:entr}
The scaling forms in Eqs.~(\ref{eq:scas2}) and (\ref{eq:scal2}) allow
us to determine the universal temperature dependence of various
thermodynamic quantities which may be measured in experiments.
Let us start with the entropy and the specific heat.
At finite temperature $T$ the renormalization group flow stops at
$\Delta_0 \sim k_BT$ due to thermal fluctuations which prevent the
formation of even larger effective spins.
At this point, all pairs of spins in links with gaps larger than
$\Delta_0\sim k_BT$ form large effective spins.
Since the distribution of gaps is broad, the interaction energies
between the effective spins are typically much smaller than
$\Delta_0\sim k_BT$, and each large spin moves essentially independently.
The entropy per unit length is hence
\begin{equation}
\frac{\sigma (T,H=0)}{L}
\propto\frac{k_B\ln(2\langle S_{\rm eff}\rangle+1)}{n}
\propto T^{2\alpha}|\ln T|
\label{eq:ent1}
\end{equation}
in the scaling regime where $\langle S_{\rm eff}\rangle\gg1$.
Note that the assumption of independent effective spins leads to an
overestimate of the entropy.
From the relation $C(T)=T\frac{d\sigma}{dT}$ follows the
specific heat per unit length 
\begin{equation}
\frac{C(T,H=0)}{L} \propto T^{2\alpha}|\ln T| \ \ .
\label{eq:spe1}
\end{equation}
This is qualitatively different than in the random spin-$\frac{1}{2}$
antiferromagnet where $\sigma (T)\propto|\ln T|^{-2}$.
This is also in contrast to the uniform 1D antiferromagnet where
$\sigma_{AF}(T)\propto T$ and the uniform ferromagnet,
$\sigma_{FM}(T)\propto\sqrt{T}$.
The fact that both the entropy and the specific heat of the
random-exchange spin chains go to zero with a rather small power
reflects the presence of large number of uncorrelated spin degrees
of freedom at low temperature. 

\subsection{Static magnetic susceptibility}
By analogous arguments, the essentially uncorrelated
large effective spins give a Curie-like contribution to
the magnetic susceptibility per unit length 
\begin{equation}
\frac{\chi}{L} =
\frac{\mu^2}{3k_BT}\frac{\langle S^2_{\rm eff}\rangle}{n}=
\frac{c}{T}\frac{\Delta_0^{-2\alpha}}{\Delta_0^{-2\alpha}}=
\frac{c}{T} \ \ .
\label{eq:susc1}
\end{equation}
The $T^{-1}$ Curie behavior is usually a signature of uncorrelated
spins.
We emphasize that this is {\it not} the case for the random spin
chain.
Rather, most of the original spins are {\it strongly} correlated, and
the Curie-like temperature dependence follows from the scaling
relation $n\sim\langle S^2_{\rm eff}\rangle$, Eq.~(\ref{eq:rawa2}). 
The Curie constant $c$ can be calculated in terms of the original spin 
distributions as follows. As discussed in Sec.~\ref{sec:scal}, the
magnitude of the effective spin representing a segment of $n$ frozen
spins is given by the sum
\begin{equation}
S^2_{\rm eff} = \left( \sum_{i=1}^n\delta_iS_i\right)^2 \ \ ,
\label{eq:seff}
\end{equation}
where the staggering factor $\delta_i$ is defined by
\begin{equation}
\delta_{i+1} = -\delta_i{\rm sgn}(J_i) \quad ; \quad \delta_1 = 1 \ \ ,
\label{eq:deld}
\end{equation}
and $ S_i >0 $ is the spin size of the elementary spin at site $ i $.
Averaging over the (initial) disorder, we obtain
\begin{equation}
\langle S^2_{\rm eff}\rangle  =  \langle\left(
\sum_{i=1}^n\delta_iS_i\right)^2\rangle = 
 n\langle S_i^2 \rangle
+\langle S_i \rangle^2\sum_{i\neq j}\langle\delta_i\delta_j\rangle  \ \ .
\end{equation}
Defining $p$ as the probability for a bond to be ferromagnetic and using
$\langle \delta_i\delta_j\rangle = (2p-1)^{|i-j|}$ we get
\begin{equation}
\langle S^2_{\rm eff}\rangle = n\left[ \langle S^2_i\rangle
  +\frac{2p-1}{1-p} 
\langle S_i \rangle^2+{\cal O}\left(\frac{1}{n}\right)\right] \ \ ,
\label{eq:ranw}
\end{equation}
and in the limit of small $T$ (large $n$) we find the Curie constant
to be
\begin{equation}
c = \frac{\mu^2}{3k_B}\left[ \langle S^2_i \rangle+\frac{2p-1}{1-p}
\langle S_i\rangle^2\right] \ \ .
\label{eq:curc}
\end{equation}
This result coincides with the low-temperature susceptibility we
obtain for the analogous classical spin chain.\cite{furu1}
Note that this low-temperature Curie constant is in general different
from the high-temperature value
\begin{equation}
\tilde{c} = \frac{\mu^2}{3 k_B} \langle S_i (S_i +1) \rangle \ \ .
\end{equation}
It follows that $ c = \tilde{c} $ only if
\begin{equation}
\frac{1-p}{2p-1} = \langle S_i \rangle \ \ .
\end{equation}
Thus, in the random spin chain we expect the
magnetic susceptibility to cross over from one Curie-like behavior at
high temperature ($T\geq J/k_B$) to a different Curie-like regime at
low-temperature ($T\leq J/k_B$).

\subsection{Magnetization at finite $H$}
In a finite magnetic field $H$ and at finite temperature $T$ the
RG flow is interrupted either by the thermal energy
$k_BT$ or by the magnetic Zeeman energy
$E_{ZM}=\mu\langle S_{\rm eff}\rangle H$.
If $k_BT\geq E_{ZM}$ the chain is dominated by thermal fluctuations
and the magnetization is given by $\chi H$.
If $k_BT\leq E_{ZM}$ the magnetic field drives the system away from
the fixed point of zero magnetic field into a state of aligned
effective spins where the magnetization eventually saturates.
In this regime a non-zero magnetic field starts to align the effective 
spins at an energy scale $\Delta_0\sim\mu\langle S_{\rm eff}\rangle H$.
With above scaling properties this means 
$\Delta_0\sim H^{1/1+\alpha}$, so that the saturated magnetization per 
unit length becomes
$M/L\sim\mu\langle S_{\rm eff}\rangle /n\sim H^{\alpha/1+\alpha}$.
The condition that the chain is not yet dominated by thermal
fluctuations is $k_BT<\Delta_0\sim H^{1/1+\alpha}$.
Summarizing these arguments, we get  
\begin{equation}
\frac{M(T,H)}{L} \propto
\cases{
       H^\frac{\alpha}{1+\alpha} & : $T^{1+\alpha} \ll bH$ \cr
       H/T  & : $T^{1+\alpha} \gg bH$, \cr}
\label{eq:mag1}
\end{equation}
where $b$ is a dimensionful non-universal constant. Similarly the entropy
goes rapidly to zero at $T^{1+\alpha}\approx bH$ when the magnetic field
starts to align the spins.

\section{Comments on the RSRG scheme}
\label{sec:comm}
In this section we discuss the validity of our
RSRG treatment and various approximations we used.
As we have seen in the previous sections, the formation of an effective
spin yields new interactions among the remaining spins.
These interactions were calculated perturbatively, where the (average) 
perturbation parameter is $\varepsilon =\Delta_t/\Delta_0$. 
To first order in $\varepsilon$, only nearest neighbor
Heisenberg terms are induced, and the functional form of the Hamiltonian
(\ref{eq:ham1}) is preserved in the RSRG transformation.
However, the terms in (\ref{eq:ham1}) are not the only
ones allowed by the symmetry, and in general we expect more
complicated isotropic interactions to appear if higher order
corrections in $\varepsilon$ are included.
In the original MDH RSRG, $\Delta_t/\Delta_0\rightarrow 0$ as the
fixed point is approached,\cite{fish1} and the perturbative treatment
becomes exact.
In our case $\varepsilon$ stabilizes at a finite value around $0.2$ at
the fixed point.
Thus we have to analyze here to what extent higher
order terms can change our results.

The basic assumptions of the RSRG scheme are that
the two spins which are most strongly coupled to each other form one
effective spin, and that any breaking up of this spin pair
involves such a large energy that we can regard the effective spin
as a rigid object.
There are two criteria for these assumptions to be valid.
First, the energy cost for breaking up the strongest spin pair, i.e.,
the energy gap $\Delta$, must be much larger than the energy available
in neighboring spin pairs.
Second, non-nearest neighbor couplings have to fall off
sufficiently rapidly with distance, so that many weak couplings cannot 
accumulate sufficient strength to break up the spin pair.
The second criterion is essentially
equivalent to the absence of strong frustration in the system.
Below, we argue that higher order terms do not lead to violation of any
of these criteria, and hence that they do not qualitatively change
any of our conclusions.
In particular, the relations (\ref{eq:exres}) between the scaling
exponents, and hence the scaling forms (\ref{eq:scas2}) and
(\ref{eq:scal2}), are still correct.
The only impact higher order terms have, is to modify the expressions in
Eq.~(\ref{eq:traf}) for the renormalized gaps, thereby slightly changing
the average ratio $a$ in Eq.~(\ref{eq:reaa}) and the scaling exponent
$\alpha$.
The low-temperature forms of the thermodynamic quantities derived
in Sec.~\ref{sec:ther} are valid even though the actual value of the
exponent $\alpha$ may shift slightly.
In particular, it is important to note that the
Curie-like form of the magnetic susceptibility does not involve $\alpha$.

\subsection{Higher order contributions}
\label{sec:hior}
We consider effective interactions between spins which are
separated by $d$ (effective) lattice spacings.
In the RG transformation these are generated by second (and higher)
order terms in $\varepsilon$, and physically they represent
interactions mediated by excitations within the locked effective
spins.
These terms will hence appear even if they are absent in the original
Hamiltonian.
However, the interactions generated in
this way fall off exponentially with distance $d$ as
\begin{equation}
\frac{\Delta(d)}{\Delta_0} \sim \varepsilon^d \ \ ,
\end{equation}
as can be seen from the following argument.
For given energy scale $ \Delta_0 $, consider the
strongest bond with spins $ S_L $ and $ S_R $ which are coupled via
long-range interactions to the spins $ S_A $ and $ S_B $,
respectively, with the gaps $ \Delta_{AL}/\Delta_0 \sim
\varepsilon^{d_{AL}} $ and $ \Delta_{RB}/\Delta_0 \sim
\varepsilon^{d_{RB}} $ (see Fig.~7).
The induced interaction between
$ S_A $ and $ S_B $ due to second-order terms is
\begin{equation}
\frac{\Delta_{AB}}{\Delta_0} \sim \frac{\Delta_{AL}
  \Delta_{RB}}{\Delta^2_0} \sim \varepsilon^{d_{AL} + d_{RB}}\ \ .
\end{equation}
Therefore the non-nearest neighbor interactions decay
exponentially with $ \varepsilon \approx 0.2 $ at the fixed point
and cannot lead to frustration effects. Consequently, it is justified
to restrict our consideration to the dominating nearest-neighbor
interactions only. The only remaining isotropic
interactions are higher order nearest neighbor spin terms
$({\bf S}_i\cdot{\bf S}_{i+1})^m$. These terms are local and could
in principle be included when calculating the lowest energy spin
multiplet and the gap to the first excited state in a link.
These higher power spin terms are also higher order in $\varepsilon$
and are unlikely to become large enough to change the
low-energy spectra of the strongest link qualitatively.

\subsection{Three-spin decimation}
\label{sec:impa}
For some particular combinations of spins and couplings, the
renormalized gap becomes larger than the gap
just removed, $\widetilde{\Delta}>\Delta_0$.\cite{foot2}
In this section we argue that even in these cases it is justified to
use the RG transformation outlined in Sec.~\ref{sec:trgs2}.

For $\widetilde{\Delta}>\Delta_0$ a more correct
procedure would be to solve the three-spin problem
involving the two spins on the strongest link and the spin on the
link with $\widetilde{\Delta}>\Delta_0$.
We would represent the ground state multiplet of the three-spin system 
with one effective spin $\widetilde{\bf S}$ and finally calculate the
effective couplings between $\widetilde{\bf S}$ and its neighbors
(Fig.~\ref{fig:laga}b).
Here we claim that we can obtain essentially the same result using our
RSRG scheme (Fig.~\ref{fig:laga}a).
In the first step the strongest link is replaced by
an effective spin ${\bf S}^\prime$, and the gaps $\Delta_1$ and
$\Delta_2$ are renormalized.
The renormalized gap $\widetilde{\Delta}_2>\Delta_0$ by assumption
immediately becomes the largest gap in the chain so that, in the next
step, the link $\{\widetilde{\Delta}_2,{S}^\prime,S_2\}$ is replaced
by an effective spin of size
$\widetilde{S}=|{S}^\prime \pm S_2|=|S_L\pm S_R\pm S_2|$.
In this process the gaps $\Delta_3$ and $\Delta_1^\prime$ are
renormalized.
The size of the effective spin $\widetilde{S}$ in Fig.~\ref{fig:laga}a
is given by the absolute value of the (vector) sum of the spins ${\bf
S}_L$, ${\bf S}_R$, and ${\bf S}_2$ parallel or antiparallel according
to the sign of the couplings.
This is the same spin we expect for the ground state multiplet of the
three-spin system, i.e., $\widetilde{\bf S}$ in Fig.~\ref{fig:laga}b.
Similarly we get the sign of the couplings between $\widetilde{\bf S}$
and its neighbors in Fig.~\ref{fig:laga}a from (\ref{eq:traf}) by
aligning the spins according to the signs of the couplings and by
comparing the direction of the spins ${\bf S}_1$ and ${\bf S}_3$ with
the direction of the effective spin $\widetilde{\bf S}$.
The signs of the couplings obtained in this way agree with what one expects
for the signs of the corresponding couplings in the three-spin treatment
in Fig.~\ref{fig:laga}b.
Therefore there can only be a difference in the
renormalized coupling strengths between the twice-two-spin and
three-spin decimation scheme.
In both ways, however, the effective couplings
$\widetilde{\Delta}_1$ and $\widetilde{\Delta}_3$ are
proportional to $\Delta_1$ and $\Delta_3$, respectively (and
independent of the magnitude of the unphysically large gap
$\widetilde{\Delta}_2$ in the RG treatment).
Therefore the discrepancy is minor and will not cause any qualitative
difference.

\section{Conclusions and summary}
\label{sec:conc}
We have studied spin chains with random couplings and random
spin sizes by means of a real-space RG scheme that successively
replaces strongly correlated spin-pairs by effective spins.
The RG transformation preserves the functional form of the Hamiltonian
but changes the probability distribution of the links
(couplings and spins).
This procedure generates interactions among the remaining spins. 
For low enough energies the probability distribution of links
acquire a scaling form, and for not too singular initial distributions
of gaps (so that $\Delta^{y_c}P(\Delta)$ is regular, $y_c\approx 0.7$)
the fixed-point distribution is universal. 
From a random walk picture for the formation of the large effective
spins, we argue for a relation between the scaling exponents of
length and the average spin size, which together with other
considerations reduces the number of independent scaling exponents to one.
This is confirmed in numerical simulations of random spin chains, and
in the universal regime, we numerically determine the remaining
independent scaling exponent to $\alpha=0.22\pm0.01$.

At low energies (low temperatures) the random spin chain is
characterized by large effective spins which interact weakly with
their nearest neighbors.
As temperature is further lowered, the average size of the effective
spins increases as $T^{-\alpha}$ while the average distance between
two effective spins (in units of the original lattice constant)
increases as $T^{-2\alpha}$.
This regime is also characterized by universal temperature dependence
of thermodynamic quantities.

The slow approach to the fixed point (our numerical simulations
indicate a crossover region of more than five orders of magnitude for
reasonable starting configurations) suggests that the true scaling
regime may be hard to reach in experiments.
However, the formation of large effective spins occurs at considerably
higher energy scale, and even if the scaling exponent $\alpha$ may not
have stabilized to its fixed-point value, the distribution of links is 
roughly like the fixed-point distribution.
The clearest signal of the formation of large effective spins is
perhaps the Curie-like temperature dependence of the uniform magnetic
susceptibility, $\chi (T)\propto 1/T$, in a temperature regime
$k_BT\leq J$ ($J$ being the typical exchange interaction in the
initial spin chain).
Also, since the $1/T$ dependence in $\chi$ emerges
before the distribution of links approaches the fixed point, the
Curie-like susceptibility should be easier to address experimentally
than other thermodynamic quantities which may develop scaling behavior
only at inaccessibly low temperatures.

When the scaling regime is realized, the most straightforward way to
measure the exponent $\alpha$, and hence the rate at which spin
degrees of freedom freeze out is through the specific heat,
$C(T)\propto T^{2\alpha}|\ln T|$. 
An alternative approach which avoids the difficulties connected with
measuring small heat transfers at low temperatures is to lower the
temperature in the presence of a magnetic field until the
magnetization saturates.
The scaling exponent $\alpha$ may then be deduced from the predicted
field dependence of the saturated magnetization,
$M_{\rm sat}(H)\propto H^{\frac{\alpha}{1+\alpha}}$.

The low-energy physics of the random spin chain studied in this paper
is very different from that of uniform spin systems as well as
spin-$\frac12$ random bond antiferromagnetic chains studied
in Refs.~\onlinecite{ma1,hirs1,hida2,fish1,hyma1,hyma2}.
In the RAF the ground state is a {\it random singlet phase}
(RSP)\cite{bhat1} where each spin forms a singlet with another spin
which may be located far away.
In the RSP the coupling between two spins that have survived down to
some energy scale $\Delta_0$ is mediated by virtually exciting all
intermediate singlets, leading to effective couplings that decreases
exponentially with length, $J\propto{\rm exp}(-\sqrt{n})$.
By inverting this relation it follows that length scales
logarithmically with  the energy, $n\propto |\ln \Delta_0|^2$.
For the RSP, arguments analogous to those in Sec.~\ref{sec:ther}
lead to entropy $\sigma (T,H=0)/L\propto |\ln T|^{-2}$ and magnetic 
susceptibility $\chi (T)/L\propto T^{-1}|\ln T|^{-2}$.\cite{fish1}
In our terminology this corresponds to $\alpha = \delta = 0$ up to
logarithmic corrections. This is consistent with the interpretation of 
$\alpha$ in terms of the average renormalization factor $a$,
Eq.~(\ref{eq:reaa}).
Indeed, in the case of the RAF the perturbative parameter
$\varepsilon=\Delta_t/\Delta_0$ and hence also $a$ goes
to zero\cite{foot4} near the fixed point, implying $\alpha =0$.
Clearly the RSP is distinct from the large-spin phase (LSP).

As pointed out in Ref.~\onlinecite{hyma1},
a third possible state of random spin chains at low temperatures is
the random dimer solid (RDS), which is easily understood within the
MDH RG picture.
Assuming that e.g., odd links are on average slightly stronger than
even links in an antiferromagnetic spin-$\frac12$ chain, 
the even links are correspondingly more likely to be removed.
Since the removal of an even (odd) link leaves a renormalized odd (even)
link behind, odd links are on average renormalized more frequently than
even ones, and the separation in energy scale between even and odd links
becomes more and more pronounced until all the singlets are on even links,
the {\it random dimer solid}. Unlike the RSP and the LSP, thermodynamic
quantities in the RDS show {\it non-universal} temperature
dependence.\cite{hyma1}

The existence of various low-temperature fixed-points raises the
question of stability of the various phases.
As is expected from the discussion above, the RSP is unstable towards
dimerization.\cite{hyma1}
In contrast, the large-spin fixed point is stable towards
dimerization.
Unlike the RSP in which spins are always removed in pairs, the RG
transformation in general removes only one spin (i.e., replaces two
spins with one effective spin).
Hence odd links are turned into even ones and vice versa so that
dimerization is irrelevant at the large-spin fixed point.
From the discussion in Sec.~\ref{sec:scal} it is also clear that both
the RSP and the RDS, which are singlet ground states, are unstable
towards a small fraction of randomly distributed ferromagnetic bonds
and/or large spins ($S>1/2$).
In both cases we expect the spin of the ground state to
scale with length as $S^2\sim L$ consistent with the fixed point
studied in this paper.
This has been confirmed in numerical simulations of random
antiferromagnetic chains close to the random-singlet fixed point with
$5\%$ ferromagnetic bonds (chain E in Tab.~\ref{tab:chains}) or with
$4\%$ of the spins $S=1$ (chain G).
In all cases studied the chains first approach the fixed-point
distribution of the RSP by forming singlets through the removal of
$S=1/2$ spins and AF links.
However, as the density of higher spins and/or FM links
increases, larger effective spins start to form and the distribution of
links crosses over to the fixed-point distribution of the LSP.
Which fixed point (either universal or non-universal) is eventually
approached depends crucially on the initial distribution.
This is because the singlet formation is very efficient in decreasing
the effective couplings and quickly builds up a singular distribution
of gaps.
Hence, as long as only singlets are formed the distribution rapidly
approaches the very singular random singlet fixed-point distribution.
The degree of the singularity in the gap-distribution at the point
where the density of $S\neq 1/2$ and ferromagnetic couplings becomes
substantial, determines the behavior of the spin chain.
If the singularity generated at this point is less than
$P(\Delta )\sim\Delta^{-y_c}$ the chain will flow to the universal
large-spin fixed point, while for stronger singularity the
chain approaches one of the non-universal large-spin fixed points.
This opens up an interesting possibility to access the non-universal
fixed points in experiments by starting with a RAF with a properly
chosen small fraction of FM bonds.

An interesting open question is what happens away from the Heisenberg point.
Fisher extended the MDH RG to anisotropic
antiferromagnetic spin-$\frac12$ chains to show that also in the XY-regime
($J^z_i<J^x_i=J^y_i$) they flow to a random singlet fixed point, at least
for broad enough initial distributions.\cite{fish1}
The inclusion of anisotropy in the generalized MDH RG and its impact on
the large-spin fixed point are interesting problems which we leave for
future research.

\acknowledgements
We would like to thank H.-C.~zur Loye and T.~N.~Nguyen for stimulating our
interest in the problem, and N.~Nagaosa and T.~K.~Ng for helpful discussions.
We are also grateful for the financial support by the Swedish Natural
Science Research Council (E.W.), by Swiss Nationalfonds (M.S.,
No.~8220--037229), and by the MIT Science Partnership Fund.
The work at MIT was supported primarily by the MRSEC Program of the National
Science Foundation under Award Number DMR-9400334.

\appendix

\section{Derivation of the effective Hamiltonian}
\label{sec:appgap}
In this appendix we give a brief derivation of the effective couplings,
Eqs.~(7) and (9).

\subsection{First-order perturbation theory}
\label{sec:appgapA}
Consider the four-spin Hamiltonian
\begin{equation}
{\cal H} = {\cal H}_0+{\cal H}_I
\label{eq:ham4}
\end{equation}
where
\begin{mathletters}
\label{eq:ham5}
\begin{eqnarray}
{\cal H}_0 & = & J_0{\bf S}_L\cdot{\bf S}_R \label{eq:ham5a} \\
{\cal H}_I & = & J_1{\bf S}_1\cdot{\bf S}_L+J_2{\bf S}_R\cdot{\bf S}_2 \ \ .
\label{eq:ham5b}
\end{eqnarray} 
\end{mathletters}\noindent
We treat ${\cal H}_I$ as a perturbation to
${\cal H}_0$.
In the space of the degenerate ground states of ${\cal H}_0$, the spins 
${\bf S}_L$ and ${\bf S}_R$ form a state of maximum (minimum) total
spin $S$ for $J_0<0$ ($J_0>0$) while the spins 
${\bf S}_1$ and ${\bf S}_2$ can point in any direction. The degenerate
ground states span the Hilbert space ${\bf H}$, the product space of
the spin spaces for ${\bf S}_1$, ${\bf S}$ and ${\bf S}_2$. Each
state 
$|m_1Mm_2\rangle=|m_1\rangle\otimes |M\rangle\otimes |m_2\rangle$ in
${\bf H}$ is labeled by the corresponding azimuthal quantum numbers $m_1$,
$M$ and $m_2$. ${\cal H}_I$ partly lifts the degeneracy and
induces an effective Hamiltonian ${\cal H}^{\rm eff}$
in ${\bf H}$. To order $J_{1,2}/J_0$ the
matrix elements of ${\cal H}^{\rm eff}$ are
\cite{land1} 
\begin{equation}
{\cal H}^{\rm eff}_{m_1Mm_2,m_1^\prime M^\prime m_2^\prime} =
\langle m_1Mm_2|{\cal H}_I|m_1^\prime M^\prime m_2^\prime\rangle \ \ .
\label{eq:pert1}
\end{equation}
We calculate ${\cal H}^{\rm eff}$ in two steps:
We first establish the operator identities (valid in ${\bf H}$)
\begin{mathletters}
\label{eq:opid1}
\begin{eqnarray}
{\bf S}_1\cdot{\bf S}_L & = & c(S_L,S_R,S)
{\bf S}_1\cdot{\bf S} 
\label{eq:opid1a} \\
{\bf S}_2\cdot{\bf S}_R & = & c(S_R,S_L,S)
{\bf S}_2\cdot{\bf S}
\label{eq:opid1b} 
\end{eqnarray}
\end{mathletters}\noindent
which, together with Eqs.~(\ref{eq:ham5b})
and (\ref{eq:pert1}), give 
\begin{equation}
{\cal H}^{\rm eff} =
\widetilde{J}_1{\bf S}_1\cdot{\bf S}
+\widetilde{J}_2{\bf S}\cdot{\bf S}_2 
\label{eq:effH2}
\end{equation}
with
\begin{mathletters}
\label{eq:effJ2}
\begin{eqnarray}
\widetilde{J}_1 & = & J_1c(S_L,S_R,{S})
\label{eq:effJ2a} \\
\widetilde{J}_2 & = & J_1c(S_R,S_L,{S}) \ \ .
\label{eq:effJ2b} 
\end{eqnarray}
\end{mathletters}\noindent
We then determine the constant $c(S_L,S_R,{S})$ by calculating a suitable
matrix element of the two operators in Eq.~(\ref{eq:opid1a}).
To establish (\ref{eq:opid1a}) we define the usual raising and
lowering operators $\hat{S}^\pm$ which act on a state $|M\rangle$ in a 
spin-$S$ multiplet in the following way,
\begin{eqnarray}
\hat{S}^- |M+1\rangle & = & A_M^S|M\rangle \nonumber \\
\hat{S}^+ |M\rangle & = & A_M^S|M+1\rangle  \ \ , \nonumber
\end{eqnarray}
where the constants $A_M^S=\sqrt{(S-M)(S+M+1)}$ assure that 
$\langle M|M\rangle =1$. Consider first the
operators $\hat{S}^+_1\hat{S}^-$ and  $\hat{S}^+_1\hat{S}^-_L$:
\begin{mathletters}
\label{eq:spsm1}
\begin{eqnarray}
\langle m_1Mm_2|\hat{S}^+_1\hat{S}^-|m_1^\prime M^\prime m_2^\prime\rangle
& = &
\delta_{m_1^\prime ,m_1-1}
\delta_{M^\prime ,M+1}
\delta_{m_2^\prime ,m_2}
A_{m_1-1}^{S_1}A_M^{{S}} \label{eq:spsm1a} \\
\langle m_1Mm_2|\hat{S}^+_1\hat{S}^-_L|m_1^\prime M^\prime m_2^\prime\rangle
& = &
\delta_{m_1^\prime ,m_1-1}
\delta_{M^\prime ,M+1}
\delta_{m_2^\prime ,m_2}
A_{m_1-1}^{S_1}\langle M|\hat{S}^-_L|M+1\rangle \ \ .
\label{eq:spsm1b}
\end{eqnarray}
\end{mathletters}\noindent
The matrix element in the right hand side of Eq.~(\ref{eq:spsm1b}) is 
\begin{eqnarray}
\langle M|\hat{S}^-_L|M+1\rangle & = & 
(A^S_{M+1}A^S_{M+2}\cdots A^S_{S-1})^{-1}
\langle M|\hat{S}^-_L(\hat{S}^-)^{S-M-1}|S\rangle \nonumber \\
& = & A^S_M(A^S_{S-1})^{-1}\langle S-1|\hat{S}^-_L|S\rangle \ \ ,
\label{eq:spsm2}
\end{eqnarray}
where we have used the fact that $\hat{S}^- = \hat{S}^-_L+\hat{S}^-_R$
commutes with $\hat{S}^-_L$.
From Eqs.~(\ref{eq:spsm1a}), (\ref{eq:spsm1b}) and (\ref{eq:spsm2}) we
see that
\begin{equation}
\hat{S}^+_1\hat{S}^-_L = (A^S_{S-1})^{-1}\langle S-1|\hat{S}^-_L|S\rangle
\hat{S}^+_1\hat{S}^- 
\label{eq:smsp5}
\end{equation}
for every combination of $m_1...M^\prime$, i.e., as an operator
identity in ${\bf H}$. 
We note that $\langle S-1|\hat{S}^-_L|S\rangle$ is a real number, and
take conjugation of Eq.~(\ref{eq:smsp5}):
\[
\hat{S}^-_1\hat{S}^+_L  =  (A^S_{S-1})^{-1}\langle S-1|\hat{S}^-_L|S\rangle
\hat{S}^-_1\hat{S}^+.
\]
We obtain the operator identity
\begin{equation}
\hat{S}^x_1\hat{S}^x_L+\hat{S}^y_1\hat{S}^y_L =
(A^S_{S-1})^{-1}\langle S-1|\hat{S}^-_L|S\rangle\left(
\hat{S}^x_1\hat{S}^x+\hat{S}^y_1\hat{S}^y\right) \ \ .
\label{eq:spsm3}
\end{equation}
Equation (\ref{eq:opid1a}) then follows from Eq.~(\ref{eq:spsm3}) and
rotational invariance.
From the relation $\langle S-1|\hat{S}^-_L|S\rangle=(A^S_{S-1})^{-1}
\langle S|\hat{S}^+\hat{S}^-_L|S\rangle=2(A^S_{S-1})^{-1}
\langle S|\hat{S}^z_L|S\rangle$ we obtain
\begin{equation}
c(S_L,S_R,S) = \displaystyle\frac{\langle S|\hat{S}^z_L|S\rangle}{S} \ \ .
\label{eq:cdet}
\end{equation}
To determine $\langle S|\hat{S}^z_L|S\rangle$ in Eq.~(\ref{eq:cdet})
we evaluate the matrix element 
$\langle S|{\bf S}_L\cdot{\bf S}|S\rangle$
in two different ways. First we use ${\bf S}={\bf S}_L+{\bf S}_R$ to get
\begin{eqnarray}
\langle S|{\bf S}_L\cdot{\bf S}|S\rangle & = &
\langle S|{\bf S}^2_L|S\rangle
+\langle S|{\bf S}_L\cdot{\bf S}_R|S\rangle \nonumber \\
& = & \frac12 \left[ S(S+1)+S_L(S_L+1)-S_R(S_R+1)\right] \ \ . 
\label{eq:way1}
\end{eqnarray}
The same matrix element can also be written as
\begin{equation}
\langle S|{\bf S}_L\cdot{\bf S}|S\rangle  = 
\langle S|\hat{S}^z_L\hat{{S}}^z|S\rangle
+\displaystyle{\frac12}\langle S|\hat{S}^-_L\hat{{S}}^+|S\rangle
+ \displaystyle{\frac12}\langle S|\hat{S}^+_L\hat{{S}}^-|S\rangle 
\ \ .
\label{eq:way21} 
\end{equation}
The first term in Eq.~(\ref{eq:way21}) equals 
$S\langle S|\hat{S}^z_L|S\rangle$, the second term vanishes since $|S\rangle$
is a highest weight state, and for the same reason we can replace the operator
$\hat{S}^+_L\hat{{S}}^-$
in the third term by the commutator
$[\hat{{S}}^+_L,\hat{{S}}^-]=2\hat{S}^z_L$ so that 
\begin{equation}
\langle S|{\bf S}_L\cdot{\bf S}|S\rangle =
(S+1)\langle S|\hat{S}^z_L|S\rangle \ \ .
\label{eq:way22}
\end{equation}
From Eqs.~(\ref{eq:way1}), (\ref{eq:way22}), and (\ref{eq:cdet}) finally
follows
\begin{equation}
c(S_L,S_R,S) = \frac{S(S+1)+S_L(S_L+1)-S_R(S_R+1)}{2S(S+1)} \ \ .
\label{eq:c123}
\end{equation}
For the three cases of interest to us, Eqs.~(\ref{eq:c123}) and
(\ref{eq:effJ2}) give
\begin{equation}
\widetilde{J}_1 = \cases{
         J_1\displaystyle\frac{S_L}{S_L+S_R} & : $J_0<0$ \cr
         J_1\displaystyle\frac{S_L+1}{S_L-S_R+1} & : $J_0>0$, $S_L>S_R$ \cr
        -J_1\displaystyle\frac{S_L}{S_R-S_L+1} & : $J_0>0$, $S_L<S_R$ \cr}
\label{eq:jeff1}
\end{equation}
and
\begin{equation}
\widetilde{J}_2 = \cases{
         J_2\displaystyle\frac{S_R}{S_L+S_R} & : $J_0<0$ \cr
        -J_2\displaystyle\frac{S_R}{S_L-S_R+1} & : $J_0>0$, $S_L>S_R$ \cr
         J_2\displaystyle\frac{S_R+1}{S_R-S_L+1} & : $J_0>0$, $S_L<S_R$ \cr}
\label{eq:jeff2}
\end{equation}
from which Eqs.~(\ref{eq:traf}) follow by using the relation between
couplings $J$ and gaps $\Delta$, Eqs.~(\ref{eq:gaps}).

\subsection{The effective coupling in the case of singlet formation: 
\label{sec:appgapB}
generalization of the MDH RG transformation}
In the case where $J_0>0$ and $S_L=S_R$, the effective spin is a singlet,
$S=0$, and the effective Hamiltonian, Eq.~(\ref{eq:pert1}), vanishes.
To get a nonzero coupling between $S_1$ and $S_2$, we have to include
second order perturbation.
In the case $S_L=S_R=\frac12$ this gives the effective couplings
used in the MDH RG. Since the first order contribution is zero, we have to
solve the second-order secular equation \cite{land1}
\begin{equation}
{\rm Det}\left( V_{gg^\prime}-E^{(2)}\delta_{gg^\prime}\right) = 0
\label{eq:secu1}
\end{equation}
with
\begin{equation}
V_{gg^\prime} = \displaystyle\sum_e
\displaystyle\frac{\langle g|{\cal H}_I|e\rangle\langle 
e|{\cal H}_I|g\rangle}{E_0-E_e} \ \ ,
\label{eq:secu2}
\end{equation}
where $|g\rangle$ is a ground state of ${\cal H}_0$, the sum is over all 
excited states $|e\rangle$ and $E_e$ is the (unperturbed) energy of
the state 
$|e\rangle$. Since there is no coupling between the outer spins if
either $J_1$ or $J_2$ is zero, only terms proportional to $J_1J_2$ in
Eq.~(\ref{eq:secu2}) contribute to the effective coupling
(the terms proportional to $J^2_1$ and $J^2_2$ give an overall energy shift
which, for our purposes, can be discarded).
Denoting the states formed by the spins $S_1$ and $S_2$ by $|A\rangle$, the
states formed by $S_R$ and $S_L$ by $|B\rangle$ and in particular the
ground state singlet by $|B=0\rangle$ we have (in obvious notation)
\begin{eqnarray}
V_{AA} & = & 2J_1J_2\displaystyle\sum_{A^\prime ,B^\prime}
\displaystyle\frac
{\langle A,0|\hat{S}^z_1\hat{S}^z_L|A^\prime ,B^\prime\rangle
 \langle A^\prime ,B^\prime|\hat{S}^z_R\hat{S}^z_2|A,0\rangle}
{-J_0\textstyle\frac12 S_{B^\prime}(S_{B^\prime} +1)} \nonumber \\
 && + \displaystyle\frac{J_1J_2}{2}
\displaystyle\sum_{A^\prime ,B^\prime}\displaystyle\frac
{\langle A,0|\hat{S}^-_1\hat{S}^+_L|A^\prime ,B^\prime\rangle
 \langle A^\prime ,B^\prime|\hat{S}^-_R\hat{S}^+_2|A,0\rangle}
{-J_0\textstyle\frac12 S_{B^\prime}(S_{B^\prime} +1)} \nonumber \\
 && + \displaystyle\frac{J_1J_2}{2}
\displaystyle\sum_{A^\prime ,B^\prime}\displaystyle\frac
{\langle A,0|\hat{S}^+_1\hat{S}^-_L|A^\prime ,B^\prime\rangle
 \langle A^\prime ,B^\prime|\hat{S}^+_R\hat{S}^-_2|A,0\rangle}
{-J_0\textstyle\frac12 S_{B^\prime}(S_{B^\prime} +1)} \ \ .
\label{eq:vele}
\end{eqnarray}
The sum over the states $|A^\prime\rangle$ can be performed separately
to give
\begin{eqnarray}
V_{AA} & = & 
-\displaystyle\frac{4J_1J_2}{J_0}\langle A|\hat{S}^z_1\hat{S}^z_2|A\rangle
\displaystyle\sum_{B^\prime}\displaystyle\frac
{\langle 0|\hat{S}^z_L|B^\prime\rangle\langle B^\prime|\hat{S}^z_R|0\rangle}
{S_{B^\prime}(S_{B^\prime} +1)} \nonumber \\
 && - \displaystyle\frac{J_1J_2}{J_0}
\langle A|\hat{S}^+_1\hat{S}^-_2|A\rangle
\displaystyle\sum_{B^\prime}\displaystyle\frac
{\langle 0|\hat{S}^-_L|B^\prime\rangle\langle B^\prime|\hat{S}^+_R|0\rangle}
{S_{B^\prime}(S_{B^\prime} +1)} \nonumber \\
 && - \displaystyle\frac{J_1J_2}{J_0}
\langle A|\hat{S}^-_1\hat{S}^+_2|A\rangle
\displaystyle\sum_{B^\prime}\displaystyle\frac
{\langle 0|\hat{S}^+_L|B^\prime\rangle\langle B^\prime|\hat{S}^-_R|0\rangle}
{S_{B^\prime}(S_{B^\prime} +1)} \ \ .
\label{eq:vel2}
\end{eqnarray}
The matrix elements in the sums get contributions only from spin-$1$
multiplets, and hence we can extract the factor 
$1/[S_{B^\prime}(S_{B^\prime}+1)]=1/2$ and perform the sum over the
complete set of states $|B^\prime\rangle$. Finally, since 
\begin{equation}
\langle 0|\hat{S}^z_L\hat{S}^z_R|0\rangle =
\frac12 \langle 0|\hat{S}^+_L\hat{S}^-_R|0\rangle
= \frac12 \langle 0|\hat{S}^-_L\hat{S}^+_R|0\rangle =
\frac13 \langle 0|{\bf S}_L\cdot{\bf S}_R|0\rangle
= -\frac13 S_L(S_L+1) 
\label{eq:vel3}
\end{equation}
we have
\begin{equation}
V_{AA} \ = \ \frac{2J_1J_2S_L(S_L+1)}{3J_0}\langle A|{\bf S}_1\cdot
{\bf S}_2|A\rangle 
\end{equation}
from which it is clear that the effective Hamiltonian in this case is
\begin{equation}
{\cal H}^{\rm eff} = \widetilde{J}{\bf S}_1\cdot{\bf S}_2
\end{equation}
with
\begin{equation}
\widetilde{J} = \displaystyle\frac{2J_1J_2S_L(S_L+1)}{3J_0} \ \ .
\end{equation}
In the special case of $S_L=S_R=1/2$ this gives
$\widetilde{J}=J_1J_2/2J_0$ in agreement with Ref.~\onlinecite{ma1}.

\section{The generalization of the MDH RG flow equations}
\label{sec:appflow}
Although in this paper we have studied the RG flow by numerical
simulations, it is of interest to show that flow equations of the form
Eqs.~(\ref{eq:flow2}) exist and can be expected to have attractive fixed
point(s) to which the probability distributions eventually flow.
Rather than deriving the explicit forms of $F_1$ and $F_2$, which
are rather complicated, we will outline how this can be done and point
out to what extent the generalized flow equations (\ref{eq:flow2})
differ from the MDH RG flow equation (\ref{eq:flow1}).

We may view one step in the RG transformation as the removal of
one link $\{\Delta_0,S_L,S_R\}$ and the change of two links,
$\{\Delta_1,S_1,S_L\}$, $\{\Delta_2,S_R,S_2\}$ $\rightarrow$ 
$\{\widetilde{\Delta}_1,S_1,{S}\}$,
$\{\widetilde{\Delta}_2,{S},S_2\}$. For an infinitely long chain we
remove a small fraction of links with gaps in the interval 
$[\Delta_0-d\Delta_0,\Delta_0]$. This is illustrated in
Fig.~\ref{fig:flow}, where, for convenience, we show
$P^A(\Delta_0;\Delta,S_L,S_R)$ and $P^F(\Delta_0;-\Delta,S_L,S_R)$ in
the same diagram.
At an energy scale $\Delta_0$, $P^A$ and $P^F$ are
non-zero only for links with $\Delta\leq\Delta_0$. We remove all links
in the thin shell $\Delta\in [\Delta_0-d\Delta_0,\Delta_0]$, which
causes a small fraction of links inside the equal-gap surface $\Delta
=\Delta_0$ to hop around.
The changes in $P^A$ and $P^F$ due to the links that move around
are of order $d\Delta_0$ so that in the limit
$d\Delta_0\rightarrow 0$ the RG flow is described by a set of first-order
differential equations
\begin{mathletters}
\label{eq:flow3}
\begin{eqnarray}
\frac{dP^A}{d\Delta_0} & = & F_1[P^A,P^F],  \\
\frac{dP^F}{d\Delta_0} & = & F_2[P^A,P^F], 
\end{eqnarray}
\end{mathletters}\noindent
where $F_1$ and $F_2$ are two (non-linear) functionals of $P^A$ and
$P^F$, whose explicit forms depend on the functions $f_n$ in
Eq.~(\ref{eq:traf}). 
If renormalized gaps were always smaller than $\Delta_0$, it would be
straightforward to write down the explicit form of the flow equations
(\ref{eq:flow2}).
Assuming no correlations between neighboring links (except for the
obvious correlation that they share one spin), we find four types of
terms in $F_1$; ({\it i}) One term proportional to
$-\delta (\Delta -\Delta_0)P^A$ that depletes the region of
links where $\Delta\in [\Delta_0-d\Delta_0,\Delta_0]$.
({\it ii}) One term that decreases $P^A$ due to links $\{\Delta,S_1,S_L\}$
that are transformed because they neighbor a link that is replaced by
an effective spin.
({\it iii}) A set of terms that increase $P^A(\Delta_0;\tilde\Delta_1,S_1,S)$
because some links are transformed into links $\{\tilde\Delta_1,S_1,S\}$.
({\it iv}) One term proportional to $P^A$ that compensates for the overall
decrease in the number of links and keeps the probability distributions
normalized. The functional $F_2$ has a similar structure.
Only the third kind of terms involve the RG functions $f_n$. 
We note that in the original MDH RG the terms ({\it ii}) and
({\it iv}) cancel. The term ({\it i}) can be omitted if the probability
distribution is defined only on the interval $[0,\Delta_0]$.
This was done in Ref.~\onlinecite{ma1} so that the only term that
appears in Eq.~(\ref{eq:flow1}) is one term of type ({\it iii}).

If some renormalized gaps become larger than $\Delta_0$, then 
before we take the limit
$d\Delta_0\rightarrow 0$ we must consider the fraction of
the transformed links that acquire
$\widetilde{\Delta}>\Delta_0$. 
As we have discussed in Sec.~VIB, in our RG scheme a
finite fraction of the links acquires a larger gap
$\widetilde{\Delta}>\Delta_0$;
in Fig.~\ref{fig:flow} this corresponds to links that jump outside the
support of $P^A$ and $P^F$.
If the unphysically strong links are not taken care of before the next
slice of links are removed, more and more links will end up outside
the $\Delta_0$ surface.
In the limit $d\Delta_0\rightarrow 0$ this will make it impossible to
define $\Delta_0$, since smoothly integrating out links in a finite
gap interval in this case generates a small but finite probability
for gaps of {\it any} strength.
This problem is remedied in the following way. 
Since in our RG scheme, no property of the renormalized links
depends on the actual value of the gap in the strongest link,
we can modify the equations (\ref{eq:egap}) so that
$\widetilde{\Delta}=\min\{\Delta f_n,\Delta_0\}$. This does not change
the discrete RG scheme since the renormalized link is removed in the
next step of the RG after which there is no information in the chain
of the value of $\widetilde{\Delta}$.
In Fig.~\ref{fig:flow}, this modification of the functions $f_n$
is illustrated by projecting strong links
back onto the surface $\Delta=\Delta_0$. The projected links are then
removed, which transforms some links in $P^A$ and $P^F$, and also
generates an even smaller fraction of links above the $\Delta_0$-surface.
These links are then projected and removed etc. 
Keeping terms of order $d\Delta_0$, the infinite
sequence of removing smaller and smaller fractions of links that 
in the previous step have jumped out of
the distribution contributes a series of smaller and smaller terms in the
shift of $P^A$ and $P^F$ that has to be summed before the limit
$d\Delta_0\rightarrow 0$ is taken. 
This resummation makes the functionals $F_{1,2}$ in (\ref{eq:flow2})
rather complicated.

\newpage

\begin{table}
\begin{center}
\begin{tabular}{cccccc}
Chain & $P_F(\Delta )$ & $P_A(\Delta )$ & $Q(S)$ & $x=N_A/N$ & length \\
\hline
A & $1-\Delta$ & $1-\Delta$ &
 $\frac{1}{20}\sum_{n=1}^{20}\delta (S-\frac{n}{2})$ &
    50\% & $10^6$ \\
B & $\Delta$ & $\Delta$ & $\frac{1}{4}\sum_{n=1}^4\delta (S-\frac{n}{2})$ &
    50\% & $10^6$ \\ 
C & $0$ & $1$ & $\frac{1}{8}\sum_{n=1}^8\delta (S-\frac{n}{2})$ &
    100\% & $10^6$ \\
D & $0.75$ & $0.25$ & $\frac{1}{4}\sum_{n=1}^4\delta (S-\frac{n}{2})$ &
    25\% & $5\cdot 10^5$ \\
E & $0.05$ & $0.2375\cdot\Delta^{-3/4}$ & $\delta (S-\frac{1}{2})$ &
    95\% & $10^6$ \\
F & $\frac{1}{2}\Delta^{-1/2}$ & $\frac{1}{2}\Delta^{-1/2}$ & 
    $\frac{1}{4}\sum_{n=1}^4\delta (S-\frac{n}{2})$ & 50\% & $10^5$ \\
G & $0$ & $\frac{1}{4}\Delta^{-3/4}$ & $0.96\delta (S-\frac{1}{2})+
    0.04\delta (S-1)$ & 100\% & $10^6$ \\
H & $\frac{2}{5}\Delta^{-3/5}$ & $\frac{2}{5}\Delta^{-3/5}$ & 
    $\frac{1}{4}\sum_{n=1}^4\delta (S-\frac{n}{2})$ & 50\% & $10^5$ \\
I & $\frac{1}{3}\Delta^{-2/3}$ & $\frac{1}{3}\Delta^{-2/3}$ & 
    $\frac{1}{4}\sum_{n=1}^4\delta (S-\frac{n}{2})$ & 50\% & $10^5$ \\
J & $\frac{2}{7}\Delta^{-5/7}$ & $\frac{2}{7}\Delta^{-5/7}$ & 
    $\frac{1}{4}\sum_{n=1}^4\delta (S-\frac{n}{2})$ & 50\% & $10^5$ \\
K & $\frac{1}{4}\Delta^{-3/4}$ & $\frac{1}{4}\Delta^{-3/4}$ & 
    $\frac{1}{4}\sum_{n=1}^4\delta (S-\frac{n}{2})$ & 50\% & $1.2\cdot10^6$ \\
L & $\frac{2}{9}\Delta^{-7/9}$ & $\frac{2}{9}\Delta^{-7/9}$ & 
    $\frac{1}{4}\sum_{n=1}^4\delta (S-\frac{n}{2})$ & 50\% & $10^5$ \\
M & $\frac{1}{5}\Delta^{-4/5}$ & $\frac{1}{5}\Delta^{-4/5}$ & 
    $\frac{1}{4}\sum_{n=1}^4\delta (S-\frac{n}{2})$ & 50\% & $10^5$ \\
N & $\frac{2}{11}\Delta^{-9/11}$ & $\frac{2}{11}\Delta^{-9/11}$ & 
    $\frac{1}{4}\sum_{n=1}^4\delta (S-\frac{n}{2})$ & 50\% & $10^5$ \\
O & $\frac{1}{8}\Delta^{-7/8}$ &  $\frac{1}{8}\Delta^{-7/8}$ & 
    $\frac{1}{4}\sum_{n=1}^4\delta (S-\frac{n}{2})$ & 50\% & $10^6$ \\
P & $\frac{1}{6}\Delta^{-5/6}$ & $\frac{1}{6}\Delta^{-5/6}$ & 
    $\frac{1}{4}\sum_{n=1}^4\delta (S-\frac{n}{2})$ & 50\% & $10^5$
\end{tabular}
\end{center}
\caption{The initial conditions for the 16 chains simulated numerically.}
\label{tab:chains}
\end{table}

\begin{figure}
\caption{Schematic pictures of the RG scheme. 
(a) The original MDH decimation.
(b) Definition of a link as two neighboring spins $S_L$ and $S_R$
and the gap $\Delta$.
(c) The generalized MDH decimation.}
\label{fig:dama} 
\end{figure}

\begin{figure}
\caption{
(a) The antiferromagnetic fixed-point distribution of spins and
gaps, $Q^A_{S\Delta}(\Delta /\Delta_0,S)=$ $\int_0^\infty
dS^\prime Q^A(\Delta /\Delta_0,S,S^\prime )$. The spins are in units of
$\langle S\rangle$ and the distribution is normalized  according to
(\protect\ref{eq:norm}). 
(b) The ferromagnetic fixed-point distribution of spins and
gaps, $Q^F_{S\Delta}(\Delta /\Delta_0,S)$, defined analogously to 
$Q^A_{S\Delta}$ in (a).
(c) The antiferromagnetic fixed-point
distributions of left and right spins, $Q^A_{SS}(S,S^\prime)=$ 
$\int_0^1 dx Q^A(x,S,S^\prime )$.
The units and normalization are as in (a).
(d) The ferromagnetic fixed-point distributions of
left and right spins, $Q^F_{SS}(S,S^\prime)$, defined analogously to 
$Q^A_{SS}$ in (c) and with
units and normalization as in (a).}
\label{fig:fpdi1}
\end{figure}

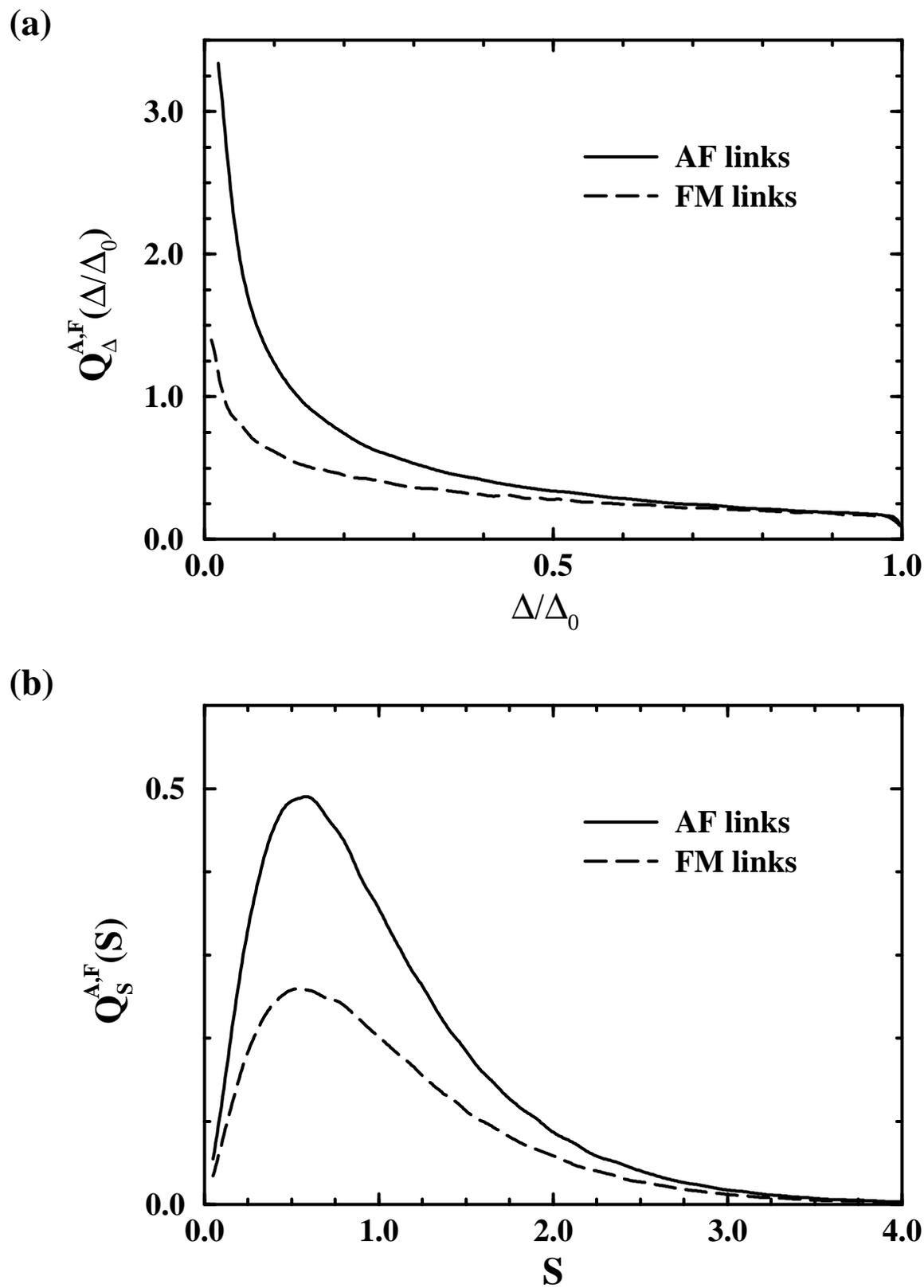
\begin{figure}
\caption{
(a) The distributions of gaps at the
fixed point,  $Q^{A,F}_{\Delta}(\Delta /\Delta_0)=$ $\int_0^\infty dS
dS^\prime Q^{A,F}(\Delta /\Delta_0,S,S^\prime )$.
(b) The distributions of spins at the
fixed point, $Q^{A,F}_{S}(S)=$ $\int_0^1 dx\int_0^\infty
dS^\prime Q^{A,F}(x,S,S^\prime )$.
The spins are in units of $\langle S\rangle$.}
\label{fig:fpdi2}
\end{figure}

\begin{figure}
\caption{
(a) The average gap $\langle\Delta\rangle$ as a
function of $\Delta_0$ in chain C in Tab.~\protect\ref{tab:chains}. 
(b) The average spin $\langle S\rangle$ and length $n$ as a function of 
$\Delta_0$ in chain C in
Tab.~\protect\ref{tab:chains}.}
\label{fig:aver}
\end{figure}

\begin{figure}
\caption{
Effective exponents as functions of $\Delta_0^{-1}$ in chain C in
Tab.~\protect\ref{tab:chains}.}
\label{fig:expo}
\end{figure}

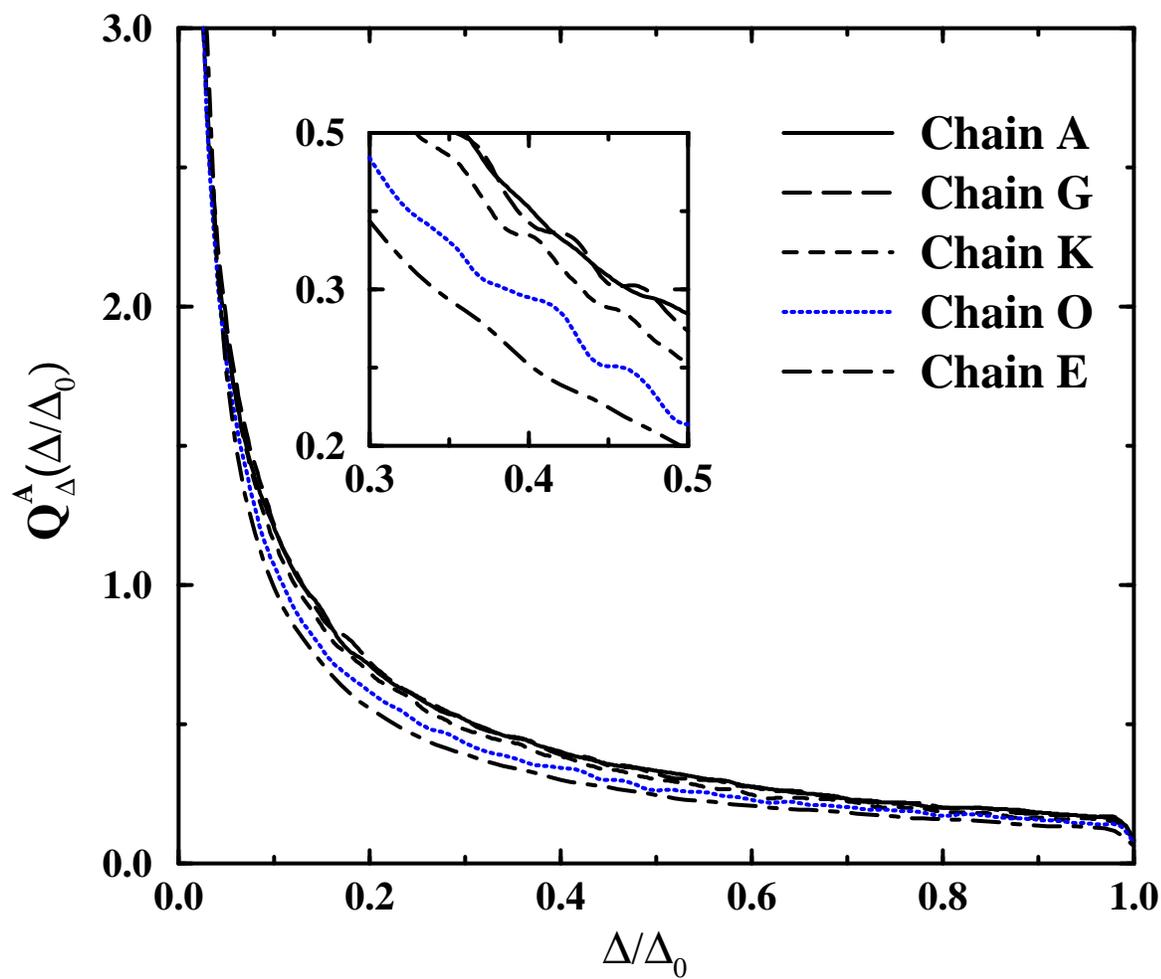
\begin{figure}
\caption{
The fixed-point distributions of AF gaps for four singular chains
(chains E, G, K and O in Tab.~\protect\ref{tab:chains}) (dotted lines).
These are to be compared with the corresponding fixed-point
distribution of regular chains (solid line)}
\label{fig:sifi}
\end{figure}

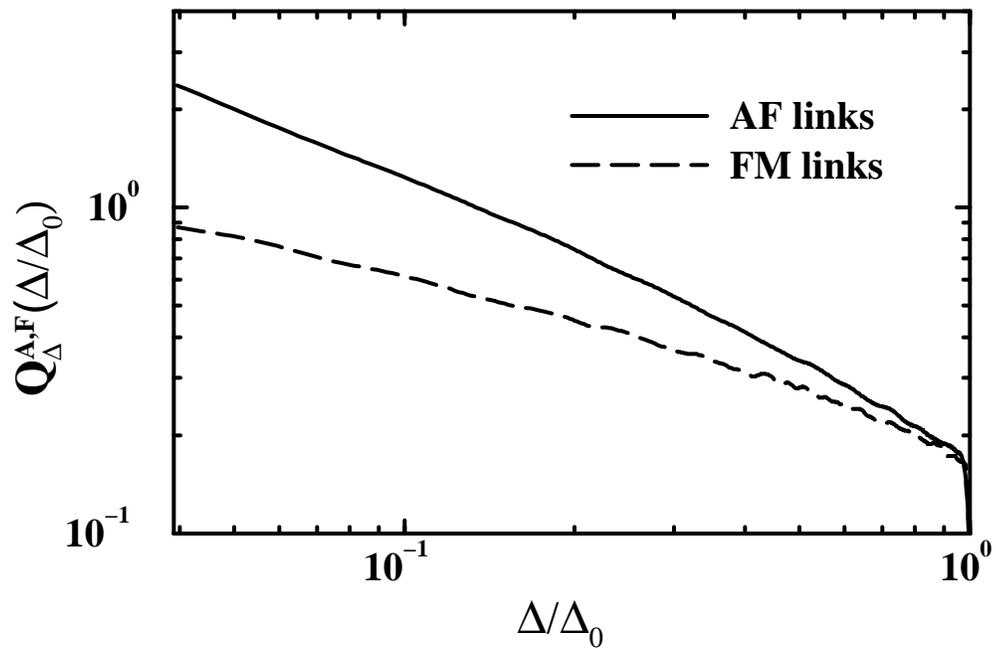
\begin{figure}
\caption{
Log-log plot of the distribution of gaps at the fixed point for
regular initial distributions.}
\label{fig:fpdilo}
\end{figure}

\begin{figure} 
\caption{
Higher order terms induce a coupling between $S_A$ and $S_B$ as $S_L$
and $S_R$ are replaced by an effective spin.}  
\label{fig:epsd} 
\end{figure}

\begin{figure}
\caption{(a) The two-step decimation of three spins used in the
numerical simulations.
(b) The three-spin decimation in one step.}
\label{fig:laga}
\end{figure}

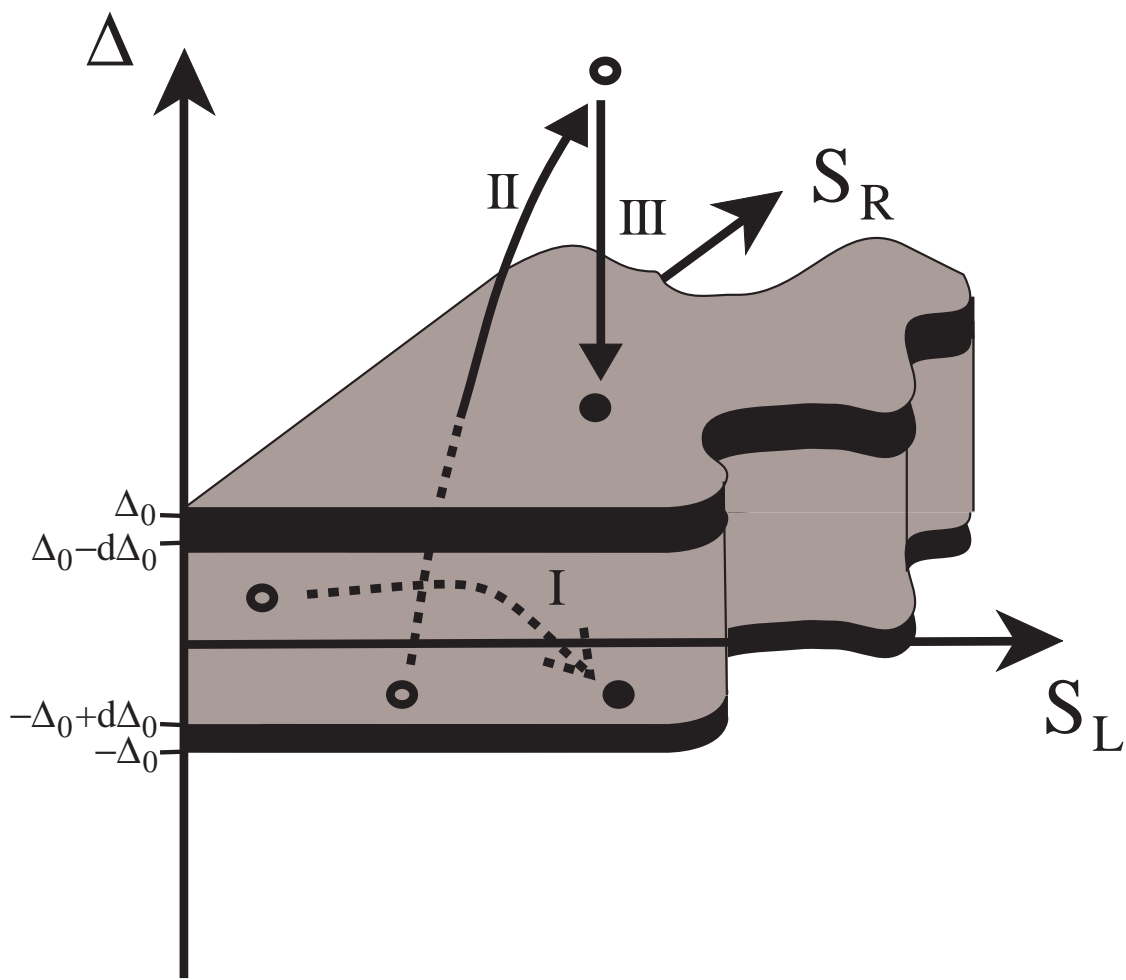
\begin{figure}
\caption{The link distributions are non-zero only for $\Delta <\Delta_0$.
Replacing gaps in the thin slice $\Delta\in [\Delta_0-d\Delta_0,\Delta_0]$
shifts a fraction of links in $P^A$ and $P^F$.}
\label{fig:flow}  
\end{figure}


\begin{references}
\bibitem{note-hyman} As pointed out by Hyman in his
thesis\protect,\cite{hyma2} this is not the only choice for the
characteristic energy scale under which one can regard two spins as a
single object.  In fact, he suggested to take $\Delta=|J|S_2(2S_1+1)$
for $S_1>S_2$, which is the gap between the ground state and highest
energy state.
\bibitem{foot2} For example, let the strongest link
be antiferromagnetic while the neighboring links are ferromagnetic and set
$S_1=10$, $S_L=2$,  $S_R=1$. Then $\widetilde{\Delta}_1$ will be larger
than $\Delta_0$ if $\Delta_1>\frac{8}{11}\Delta_0$.
\bibitem{foot4} We point out that in the random singlet phase,
$\beta\neq 1$ is allowed because the fixed-point distribution is
strictly speaking not normalizable.
\end{references}
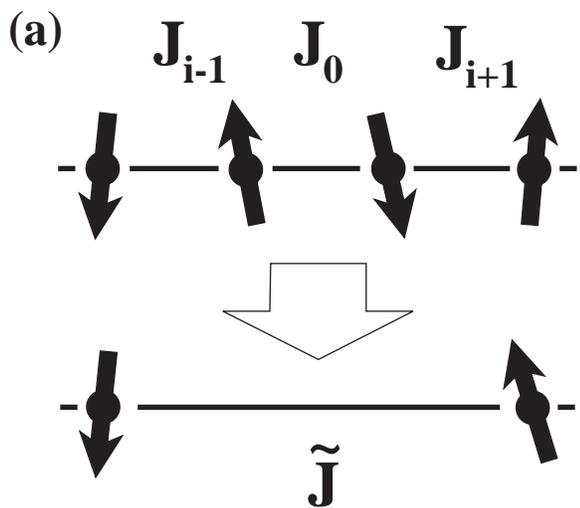
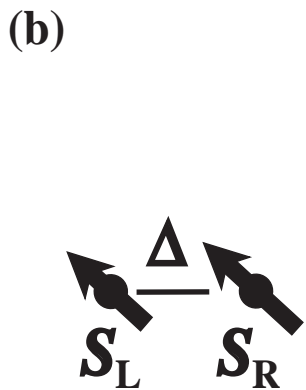
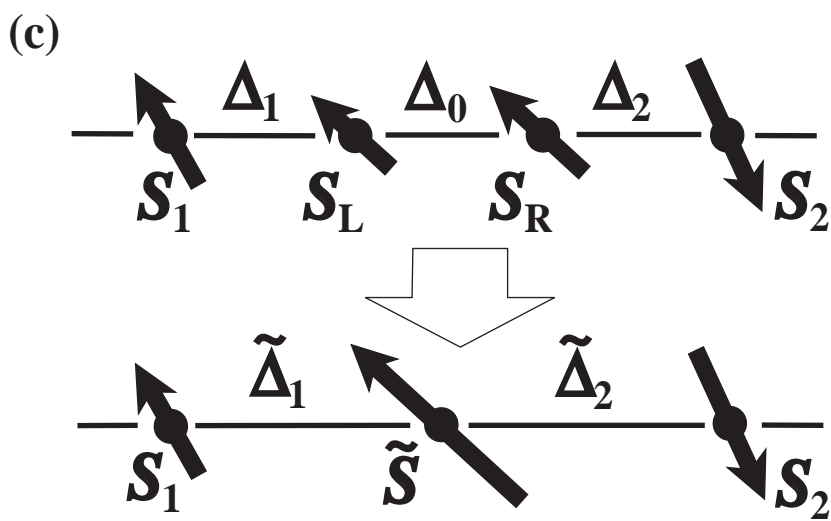
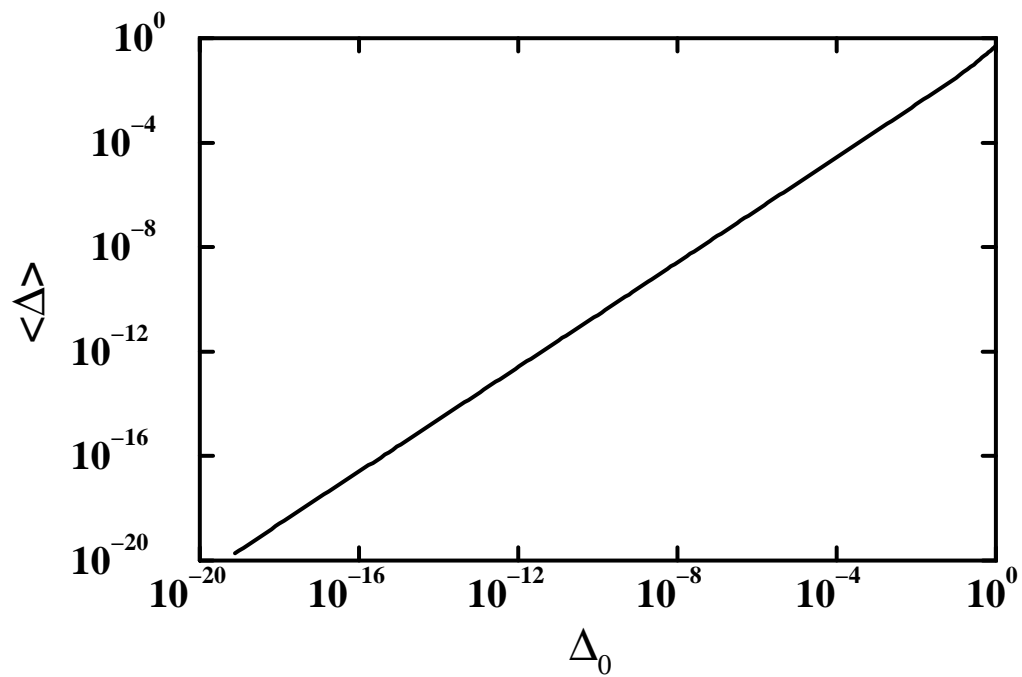
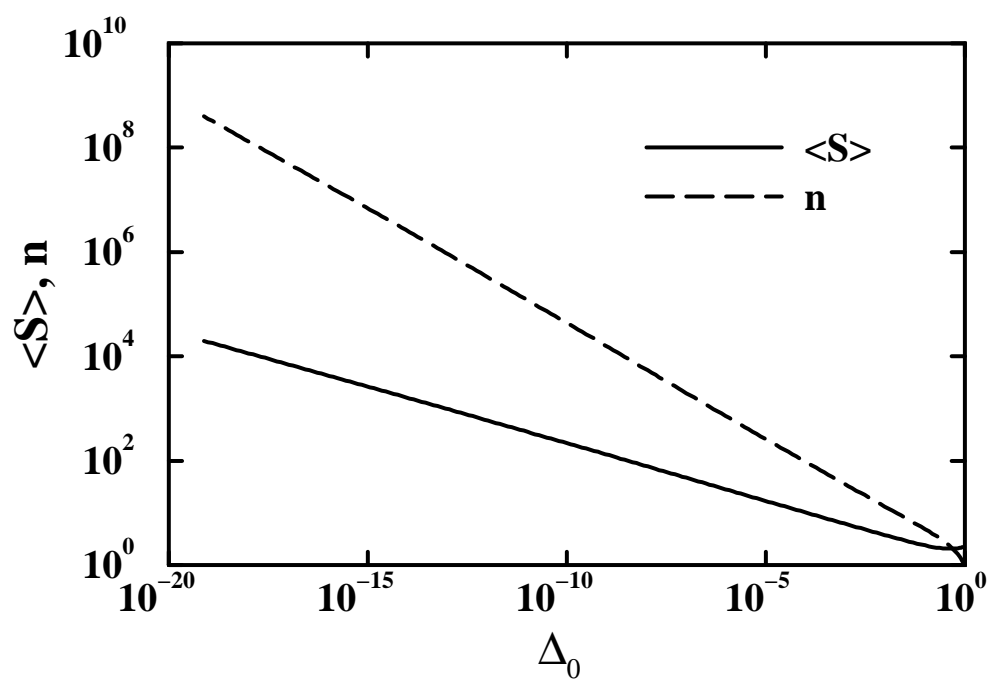
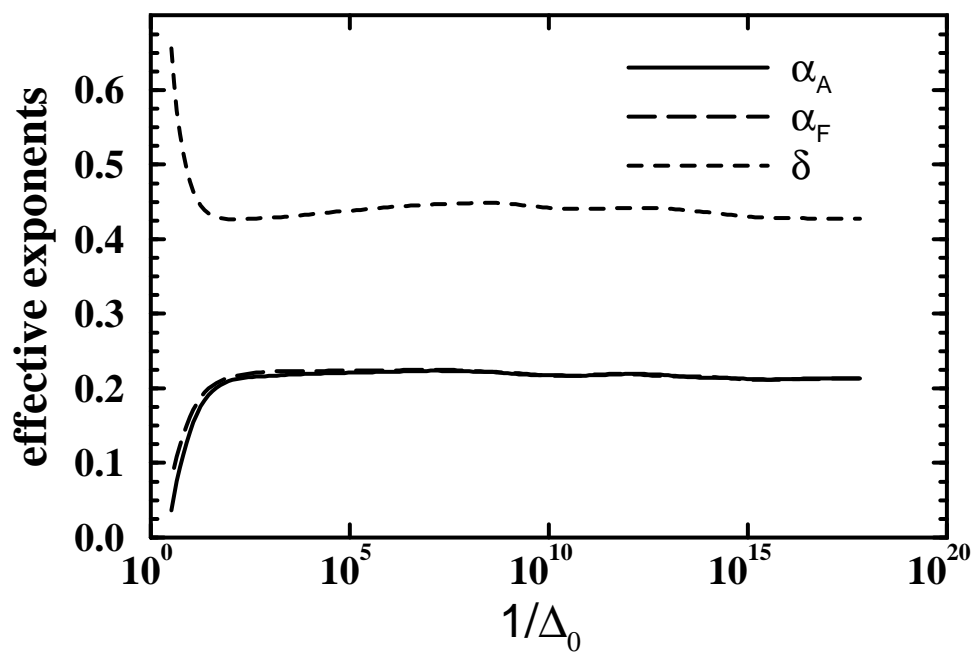
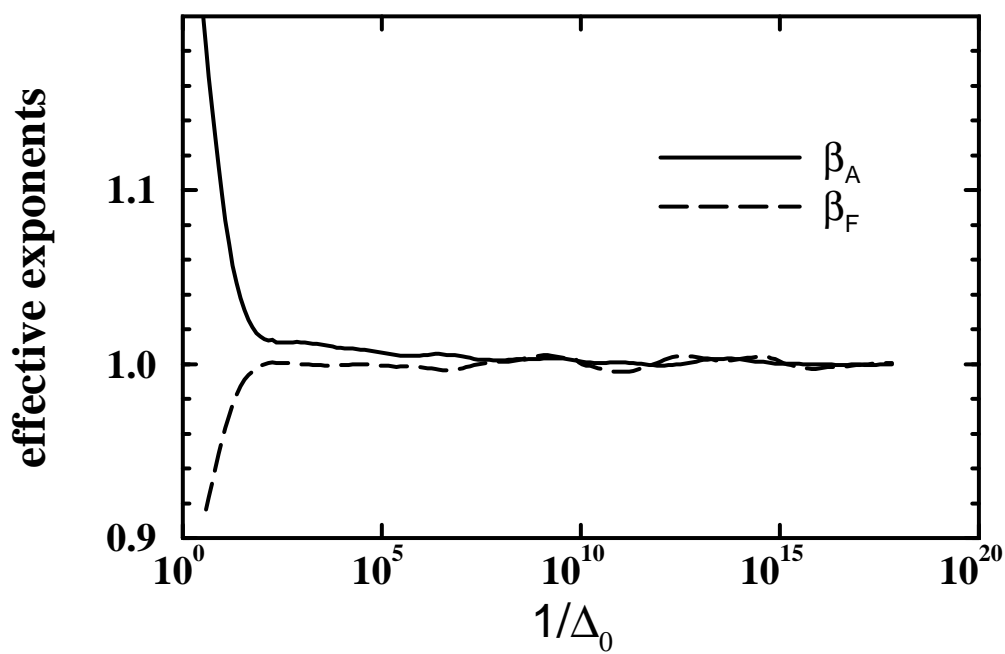
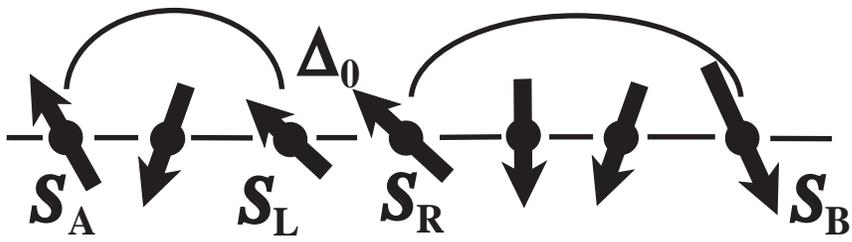
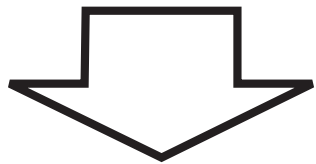
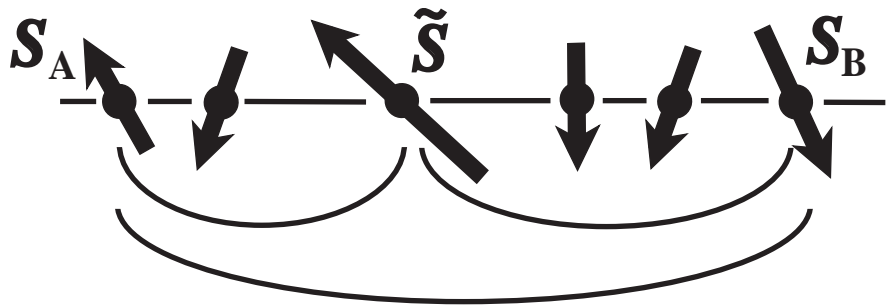
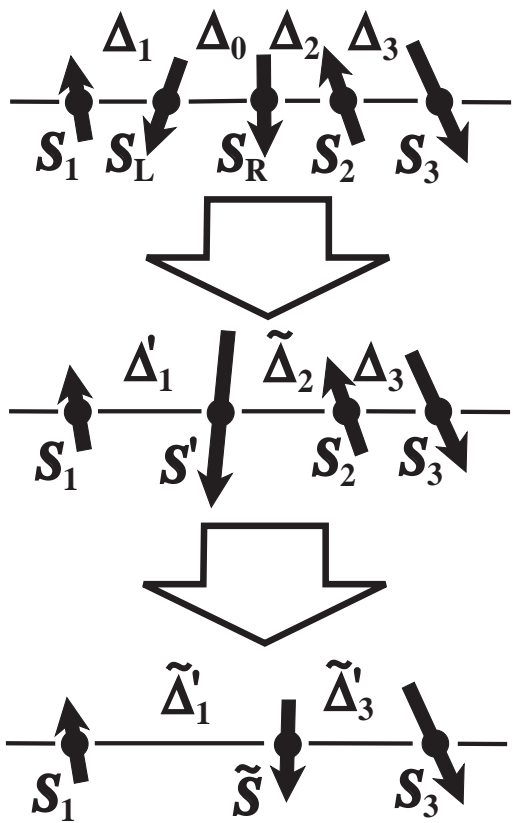
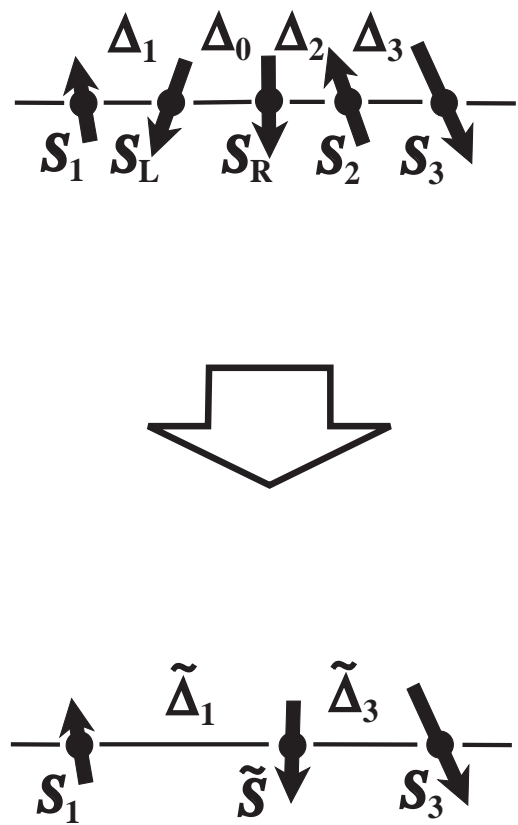
\end{document}